\documentclass[12pt,preprint]{aastex}

\newcommand{\mdot}{M$_{\odot}$ yr$^{-1}$}
\newcommand{\ldot}{L$_{\odot}$}
\newcommand{ \um}{ $\mu$m~}
\newcommand{  \ums}{ $\mu$m}

\def\kmsMpc{\ifmmode {\rm\,km\,s^{-1}\,Mpc^{-1}}\else
    ${\rm\,km\,s^{-1}\,Mpc^{-1}}$\fi}

\shorttitle{Most Luminous Dusty Galaxies}
\shortauthors{Weedman and Houck}

\begin{document}

\title{Evolution of the Most Luminous Dusty Galaxies}

\author{Daniel W. Weedman\altaffilmark{1} and James R. Houck\altaffilmark{1}}

\altaffiltext{1}{Astronomy Department, Cornell University, Ithaca, NY 14853; dweedman@isc.astro.cornell.edu}

\begin{abstract}
A summary of mid-infrared continuum luminosities arising from dust is given for very luminous galaxies, L$_{IR}$ $>$ 10$^{12}$ \ldot, with 0.005 $<$ z $<$ 3.2 containing active galactic nuclei (AGN), including 115 obscured AGN and 60 unobscured (type 1) AGN.  All sources have been observed with the $Spitzer$ Infrared Spectrograph.  Obscured AGN are defined as having optical depth $\tau$ $>$ 0.7 in the 9.7 $\mu$m silicate absorption feature (i.e. half of the continuum is absorbed) and having equivalent width of the 6.2 $\mu$m polycyclic aromatic hydrocarbon (PAH) feature $<$ 0.1 $\mu$m (to avoid sources with a significant starburst component).  Unobscured AGN are defined as those that show silicate in emission.  Luminosity $\nu$L$_{\nu}$(8 $\mu$m) for the most luminous obscured AGN is found to scale as (1+z)$^{2.6}$ to z = 2.8.  For unobscured AGN, the scaling with redshift is similar, but luminosities $\nu$L$_{\nu}$(8 $\mu$m) are approximately 3 times greater for the most luminous sources.  Using both obscured and unobscured AGN having total infrared fluxes from IRAS, empirical relations are found between $\nu$L$_{\nu}$(8 $\mu$m) and L$_{IR}$.  Combining these relations with the redshift scaling of luminosity, we conclude that the total infrared luminosities for the most luminous obscured AGN, L$_{IR}$(AGN$_{obscured}$) in \ldot, scale as log L$_{IR}$(AGN$_{obscured}$) = 12.3$\pm$0.25 + 2.6($\pm$0.3)log(1+z), and for the most luminous unobscured AGN, scale as log L$_{IR}$(AGN1) = 12.6($\pm$0.15) + 2.6($\pm$0.3)log(1+z).  We previously determined that the most luminous starbursts scale as log L$_{IR}$(SB) = 11.8$\pm$0.3 + 2.5($\pm$0.3)log(1+z), indicating that the most luminous AGN are about 10 times more luminous than the most luminous starbursts.  Results are consistent with obscured and unobscured AGN having the same total luminosities with differences arising only from orientation, such that the obscured AGN are observed through very dusty clouds which extinct about 50\% of the intrinsic luminosity at 8 $\mu$m.  Extrapolations of observable f$_{\nu}$(24 $\mu$m) to z = 6 are made using evolution results for these luminous sources.  Both obscured and unobscured AGN should be detected to z $\sim$ 6 by $Spitzer$ surveys with f$_{\nu}$(24 $\mu$m) $>$ 0.3 mJy, even without luminosity evolution for z $>$ 2.5.  By contrast, the most luminous starbursts cannot be detected for z $>$ 3, even if luminosity evolution continues beyond z = 2.5.

\end{abstract}

\section{Introduction}

Understanding the initial formation and evolution of galaxies and active galactic nuclei (AGN) within the early Universe is a major goal of observational cosmology.  While average parameters of the Universe itself and the observed clustering of galaxies are well described by conclusions from microwave background anisotropies \citep[e.g.][]{spe03}, the transformation from initial anisotropies to individual primordial galaxies during the first few billion years of the Universe (redshift z $\ga$ 2) is not understood.  When did the first galaxies assemble?  When did the first generation of stars occur?  How did these enrich the interstellar medium with metals, molecules, and dust?  When and how did the massive black holes of AGN develop?

Observations available to date \citep[e.g.][]{ric06,mad98} indicate that these formation processes maximized at  2 $\la$ z $\la$ 3 , both for AGN and star formation (hereinafter, we refer to luminous, star forming galaxies as "starbursts").   Understanding that maximum and tracing the decline of AGN and starbursts since that epoch provides fundamental insight on how these processes have evolved in the Universe. 

The capabilities of the Spitzer Space Telescope ($Spitzer$) provide powerful and independent tools to address these issues, for several reasons.  The most luminous galaxies known (L$_{IR}$ $>$ 10$^{12}$\ldot) are the Ultraluminous Infrared Galaxies (ULIRGs, e.g. Soifer et al. 1986, Sanders et al. 1988), whose luminosity arises because obscuring dust reemits luminosity absorbed from shorter wavelengths.  The total dust emission in the infrared from galaxies of all luminosities dominates the cosmic background luminosity, and this luminosity constrains the total number of galaxies and AGN \citep{cha01,lag04,lef05}. 

The extreme extinction at optical and ultraviolet wavelengths makes it impossible in many cases to observe the primary luminosity sources for AGN and starbursts.  Observing such objects in the infrared overcomes this severe extinction.  Furthermore, the wide wavelength range of the the Infrared Spectrograph on $Spitzer$ (IRS; Houck et al. 2004) makes it possible to trace the same diagnostic spectral features all the way from z = 0 to z $\sim$ 3.   

A variety of observing programs with $Spitzer$ have been dedicated to understanding the optically-faint populations of dusty extragalactic sources which are revealed in mid-infrared surveys at 24\,\um with the Multiband Imaging Photometer for $Spitzer$ (MIPS; Rieke et al. 2004) and in near-infrared surveys with the $Spitzer$ Infrared Array Camera (IRAC; Fazio et al. 2004).  A major objective has been to understand sources discovered with very large infrared to optical ratios, typically having f$_{\nu}$(24 \ums) $>$ 1 mJy and $R$ $>$ 24 mag. [i.e. f$_{\nu}$(24 \ums)/f$_{\nu}$(R) $\ga$ 1000].  

Spectroscopic observations with the IRS have shown that these dusty sources primarily fall into two categories: sources with strong absorption by the 9.7 \um silicate absorption feature, and sources with strong polycyclic aromatic hydrocarbon (PAH) emission features \citep[e.g.][]{hou05,yan07}.  In both cases, the large infrared to optical ratios are attributed to obscuration by dust, which heavily extincts the rest frame visible and ultraviolet fluxes while producing the mid-infrared continuum through reemission.  These heavily extincted sources have been termed "Dust Obscured Galaxies" \citep{dey08}.

Sources with strong PAH emission are attributed to starbursts, with the PAH arising in the photodissociation region surrounding the ionized gas produced by the young, hot stars \citep[e.g.][]{gen98,bra06}.  Sources with little PAH luminosity but strong silicate absorption are explained as active galactic nuclei (AGN) surrounded by absorbing clouds of dust \citep[e.g.][]{shi07,ima07}.  These clouds are identified primarily because of heavy absorption in the 9.7 \um silicate feature.  Dust associated with clouds in such AGN can also be seen in silicate emission, if the continuum observed from the hotter dust is not observed through cooler dust which produces the absorption \citep[e.g.][]{hao05,sch07}. 

In a previous paper \citep{wee08}, we summarized results for the most luminous dusty starbursts with 0 $<$ z $<$ 2.6 discovered by $Spitzer$ and observed with the IRS.  In the present paper, we present an analogous summary for the most luminous dusty AGN.  Because these AGN are often at z $\sim$ 2, are highly luminous, and are spatially unresolved in infrared or ground-based optical observations, they are often called quasars.  We use the term AGN because imaging of some examples with the Hubble Space Telescope shows that they have a surrounding, spatially extended component \citep{bus09}.  

In the present paper, we trace the evolution of both obscured and unobscured AGN from z = 0 to z = 2.8 using a single spectral diagnostic [$\nu$L$_{\nu}$(8 $\mu$m)], determine the comparisons of their total infrared luminosities L$_{IR}$ to $\nu$L$_{\nu}$(8 $\mu$m), and compare the AGN to analogous results for starbursts.  We also predict the f$_{\nu}$(24 \ums) for AGN and starbursts that would be observed to z = 6 based on average spectra of the most luminous examples of each class.

\section{Sample Definition and Data Analysis}

\subsection{Source Selection}

\subsubsection{IRS Spectra of Obscured Sources}

Hundreds of IRS spectra of AGN are now available within the $Spitzer$ archive, ranging from systematic observations of previously known, nearby AGN to discovery observations of sources with z $>$ 2. In this paper, we assemble data from 20 different $Spitzer$ programs to summarize a wide variety of AGN, including 115 heavily obscured sources classified by having strong absorption in the 9.7 \um silicate feature, and 60 unobscured sources (type 1 AGN) as classified by silicate emission.  

Most high redshift AGN discovered with $Spitzer$ and with redshifts measured by the IRS are obscured sources with significant silicate absorption.  This is primarily a selection effect, because the presence of strong silicate absorption provides an unambiguous spectral feature for a redshift measurement.  By contrast, AGN with no silicate absorption appear essentially as power law spectra and cannot be assigned a redshift based on the IRS spectrum. 

Our summary is intended to present an overview of a wide range of sources with IRS spectra which satisfy criteria of being AGN, either heavily obscured or unobscured.  All AGN which have been observed with $Spitzer$ are not included, especially lower luminosity, local objects.  We attempt, however, to include all of the most luminous AGN with IRS spectra at all redshifts.   The primary selection criteria are designed to locate "pure" AGN, both heavily obscured and unobscured, and to distinguish them from sources having their mid-infrared luminosity arising primarily from starbursts.  

Most of the new sources initially discovered by $Spitzer$ and having IRS followup spectra are at z $\sim$ 2, typically with f$_{\nu}$(24 $\mu$m) $\ga$ 1 mJy.  The signal to noise (S/N) of such spectra is low, so it is important to have measurement criteria which are easily applied to such faint sources.  In Figure 1, we illustrate examples of rest-frame spectra (with artificially high S/N) for the brightest sources which have been observed at z $\sim$ 2.5, scaled to flux densities actually observed for the brightest examples published so far. (These are source 90 in Table 1 (below) for the most luminous absorbed AGN, source 37 in Table 2 (below) for the most luminous type 1 AGN, and source MIPS 506 \citep{yan07} for the most luminous starburst.)  The shape of the average spectrum shown for type 1 AGN derives from the averages of the two most luminous quasars in \citet{wee09}, which are also sources 37 and 49 in Table 2.  The shape of the average spectrum of the absorbed AGN is the average of the three luminous ULIRGs discussed below in section 3.2 and shown in Figure 5.  The shape of the average spectrum of the most luminous starbursts is the average of the four most luminous starbursts in \citet{wee08}.  

In our previous discussion of starbursts with such redshifts \citep{wee08}, we measured luminosity using the peak flux density of the 7.7 $\mu$m PAH emission feature.  As can be seen in Figure 1, this is the observed peak of the spectrum for starburst sources at z $\ga$ 2.  A similar peak at $\sim$ 8 \um arises in obscured AGN for a very different reason, which is the heavy silicate absorption longward of this wavelength and the heavy absorption by ices shortward of this wavelength \citep[e.g.][]{spo04}.  The flux density f$_{\nu}$(8 \ums) of this apparent continuum peak at $\sim$ 8 \um is the best measure of luminosity of an absorbed AGN at redshifts z $\sim$ 2.  The precise rest-frame wavelength of this localized peak can range from about 7.7 \um to 8 \um, depending on the characteristics of the absorption.  Although we refer to this as the 8 \um peak, measurements that we report are the maximum flux density at $\sim$ 8 \um, regardless of the precise wavelength.

It is essential to distinguish AGN and starbursts among the faint sources discovered and observed with $Spitzer$.  Comparing spectra for the absorbed AGN and the starburst in Figure 1 indicates the importance of the PAH features at 6.2 \um and 8.6 \um for judging whether the peak at $\sim$ 8 \um is dominated by PAH 7.7 \um emission or by the dust continuum of an absorbed AGN.  Various previous studies and summaries indicate that a "pure" starburst has rest frame equivalent width (EW) of the 6.2 \um feature greater than 0.4  \ums, but that sources with EW(6.2 \ums) $<$ 0.1 \um have their mid-infrared continuum dominated by AGN luminosity \citep[e.g.][]{bra06,ima07,far07}.  

There are many intermediate cases in which both starbursts and AGN contribute to the mid-infrared luminosity; dividing such composites into AGN and starburst components is done primarily using the EW of the PAH features \citep[e.g.][]{des07,saj08,sar08}.  For individual objects, determining the relative AGN and starburst contributions is important.  In our studies, however, both for starbursts as done previously and for the AGN in the present paper, we attempt to avoid uncertainties regarding composite sources by concentrating only on sources that appear either as pure starbursts (very strong PAH) or as pure AGN (very weak PAH).  

This is an empirical approach, designed to make best use of the numerous spectra of poor S/N.  It also allows a straightforward discussion of the luminosities of AGN or of starbursts, without introducing model-dependent uncertainties arising from different methods of deconvolving composite sources. The criterion which we adopt for including sources in the present paper as AGN is that sources have a measured value of EW(6.2 \ums) $<$ 0.1 \um, or that no detection is reported for the 6.2 \um feature.

\subsubsection{Evidence that Absorbed Sources are Obscured AGN}

The type 1 AGN included in this paper are all classified using available optical spectra, so there is no ambiguity regarding the presence of an AGN.  The obscured sources suffer severe extinction in the optical, however, so optical spectra, even when available, would not necessarily reveal an obscured AGN.  Our definition of obscured AGN states that in the absence of the PAH features from starbursts, the infrared spectrum arises from dust reradiation of luminosity initially produced by an AGN.  For a source so hidden that no spectral diagnostics can escape, it is never possible to disprove the possibility that some of the luminosity arises from hidden starbursts.  There are numerous reasons to believe, however, that sources characterised by strong silicate absorption and weak PAH emission show a spectrum which arises from dust heated by an AGN and which is then absorbed by intervening, cooler dust. 

The first source for which silicate absorption was observed was the prototype Seyfert 2 galaxy NGC 1068, for which the continuum of heated dust arises from a region surrounding the AGN which is even smaller than the narrow line region \citep{tel84}.  As $Spitzer$ IRS spectra accumulated, it was found that type 2 (optically absorbed) AGN characteristically show silicate absorption, whereas type 1 (optically transparent) AGN typically show silicate emission \citep{hao07}, reinforcing the association between silicate absorption and obscured AGN.  

The most thoroughly observed obscured ULIRG is Markarian 231. This ULIRG unambiguously contains a type 1 AGN, immersed in heavy clouds that produce absorption in X-ray, ultraviolet, and visible \citep[e.g.][]{gal02}. Half of the continuum is absorbed by the silicate absorption at 9.7  \ums, so that the optical depth $\tau$ of the absorption feature is $\tau$ = 0.7 \citep{arm07}. In selecting "obscured AGN", we use a criterion that assures sufficiently significant absorption that the source can be recognized in any IRS spectrum that includes the rest-frame feature.  Our criterion requires that $\tau$ $>$ 0.7, so Markarian 23l defines the lower limit of absorption for obscured AGN which we include in the present paper. 

\citet{ima07} use measures of relative absorption in the 9.7 \um and 18 \um silicate features to show from radiative transfer arguments that the heavily absorbed ULIRGs require a concentrated source of dust heating, which arises naturally for an AGN but not for distributed starbursts.  Coupled with the weak PAH features in such sources, they conclude that these absorbed ULIRGs are dominated by AGN luminosity.

When the overall spectral energy distributions (SED) of the absorbed sources discovered by $Spitzer$ are examined, there is a distinct relation between the SED and the presence of strong silicate absorption.  The absorbed sources show power law SEDs from mid-infrared through optical wavelengths \citep[e.g.][]{wee06c,pol08}.  By contrast, selection of sources which show the stellar absorption feature at rest frame 1.6 \um invariably yields sources with PAH emission \citep{far08}.  For large samples of $Spitzer$ sources having photometry that determines the SED, the power law SEDs correlate with the presence of X-ray luminosity, leading to the conclusion that these power law sources are AGN \citep{brn06,pol07,don08,fio08}. 

For all of the above reasons, we feel it is appropriate to interpret the sources with deep silicate absorption and weak PAH emission as having the mid-infrared continuum arising because of an AGN.  As emphasized in section 2.1.1, we attempt to avoid the uncertainty of AGN+starburst composite sources by using 
only sources with with quantitatively weak PAH features. This produces a sample of sources which can be considered as the "purest" absorbed AGN, and it is this sample which is discussed below.

\subsection {$Spitzer$ IRS Samples of AGN}

Absorbed sources are chosen from various $Spitzer$ observing programs as those sources with IRS spectra which show $\tau$ $>$ 0.7 for the 9.7 \um silicate absorption feature and EW(6.2 \ums) $<$ 0.1 for the 6.2 \um PAH emission feature.  Emission sources are chosen from various programs as those sources with 9.7 \um silicate emission of any strength and EW(6.2 \ums) $<$ 0.1 for the PAH emission feature. We have not attempted to include a comprehensive list of low luminosity, local AGN observed with $Spitzer$ because our objective is to locate the most luminous sources distributed over the largest possible redshift range.  

The various observing programs by $Spitzer$ used to assemble this collection of IRS spectra of AGN are summarized below.  The numbering for the list below of these observing programs is the same as the reference codes in Table 1, which contains the sample of obscured (absorbed) AGN, and Table 2, which contains the sample of unobscured (emission) AGN. Sources discovered with $Spitzer$ MIPS 24 \um surveys derive from the NOAO Deep Wide Field Survey (NDWFS) in Bootes \citep{jan99}, the Spitzer Wide-area Infrared Extragalactic Survey (SWIRE, Lonsdale et al. 2003, 2004)\footnote{http://swire.ipac.caltech.edu/swire/astronomers}, and the Spitzer First Look Survey (FLS, Fadda et al. 2006). The sources in Tables 1 and 2 arise as follows: 
 
1. 16 absorbed and one silicate emission Ultraluminous Infrared Galaxies (ULIRGs) discovered by the Infrared Astronomical Satellite (IRAS) from 53 sources in \citet{far07}.  Except for 9 sources in \citet{arm07}, spectra in this sample are not published, so we extracted spectra from the $Spitzer$ archive for all 37 sources shown in Figure 21 of \citet{far07} as having $\tau$ $>$ 0.7 and for all 4 sources shown as having silicate emission.  Spectra were extracted as described below in section 2.3.  From these spectra, the EW(6.2 \um) was measured to determine if the sources satisfy our criteria for pure AGN.  16 absorption sources and two silicate emission sources did so, and the f$_{\nu}$(8 \ums) of these were measured from the extracted spectra.  One of the emission sources is 3C 273, included below in reference 16. 
 
2. 12 absorbed IRAS ULIRGS from 48 sources in \citet{ima07}, using those listed with $\tau$ $>$ 0.7 and EW(6.2 \ums) $<$ 0.1.  The f$_{\nu}$(8 \ums) are measured from the published spectra.  

3. 13 absorbed IRAS Faint Source Catalog (FSC) ULIRGs from 28 sources in \citet{sar08}, using those listed with $\tau$ $>$ 0.7 and EW(6.2 \ums) $<$ 0.1.  The f$_{\nu}$(8  \ums) are measured from the original spectral extractions.

4. 4 absorbed local AGN from 97 sources in \citet{shi07}. These are the 4 AGN out of 97 published spectra which have $\tau$ $>$ 0.7; the f$_{\nu}$(8 \ums) are measured from the published spectra. 

5. 3 absorbed AGN and 11 silicate emission AGN from 60 sources in \citet{wee09}, selected from the flux limited sample with f$_{\nu}$(24 $\mu$m) $>$ 10 mJy taken from the FLS and the NDWFS in Bootes. The f$_{\nu}$(8 \ums) are the published values.

6. 9 absorbed AGN with new spectra described in section 2.3.1, below, and shown in Figure 2.  Sources derive from sources with IRS spectra in the FLS and SWIRE Lockman fields having f$_{\nu}$(24 $\mu$m) $>$ 5 mJy and $R$ $>$ 20 mag; two additional sources satisfying these criteria are contained in references 5 and 12. 

7. 3 absorbed AGN from 16 sources in \citet{brn08a}, selected as X-ray sources in the NDWFS Bootes field having f$_{\nu}$(24 $\mu$m) $>$ 1 mJy and $R$ $>$ 22 mag. 

8. 2 absorbed AGN from 16 sources in \citet{far09}, selected from the SWIRE Lockman field as sources with f$_{\nu}$(70 $\mu$m) $>$ 15 mJy and $R$ $>$ 23 mag.

9. 3 absorbed AGN from 11 sources in \citet{brn08b}, selected from the Bootes field as sources with f$_{\nu}$(70 $\mu$m) $>$ 30 mJy and $R$ $>$ 22 mag. 

10. 5 absorbed AGN from 31 sources in \citet{hou05}, selected from the Bootes field as sources with f$_{\nu}$(24 $\mu$m) $\ga$ 1 mJy and $R$ $\ga$ 24 mag.  New extractions of sources in \citet{hou05} were done as described in section 2.3 below in order to determine if ambiguous cases are starburst spectra with PAH or AGN spectra with absorption.  The final sources selected are in Table 1, and the f$_{\nu}$(8 \ums) are from the new spectral extractions.  

11. 5 absorbed AGN from 18 sources in \citet{wee06a}, selected as radio sources in the FLS field with f$_{\nu}$(24 $\mu$m) $\ga$ 1 mJy and $R$ $\ga$ 24 mag.  New extractions of sources in \citet{wee06a} were done as described in section 2.3 below in order to determine if ambiguous cases are starburst spectra with PAH or AGN spectra with absorption.  The final sources selected are in Table 1, and the f$_{\nu}$(8 \ums) are from the new spectral extractions. 

12. 3 absorbed AGN from 20 sources in \citet{wee06c}, selected from the SWIRE Lockman field as sources with f$_{\nu}$(24 $\mu$m) $\ga$ 1 mJy and $R$ $\ga$ 24 mag. New extractions of sources in \citet{wee06c} were done as described in section 2.3 below in order to determine if ambiguous cases are starburst spectra with PAH or AGN spectra with absorption.  The final sources selected are in Table 1, and the f$_{\nu}$(8 \ums) are from the new spectral extractions. 

13. 7 absorbed AGN from \citet{pol08}, selected as the most luminous absorbed sources known and modeled with a dusty torus. Sources have f$_{\nu}$(24 $\mu$m) $>$ 1 mJy and $R$ $>$ 24 mag.

14. 22 absorbed AGN from 52 sources in \citet{yan07} and \citet{saj08}, selected from the FLS using the criteria  f$_{\nu}$(24 $\mu$m) $>$ 0.9 mJy, $\nu$f$_{\nu}$(24 $\mu$m)/$\nu$f$_{\nu}$(8 $\mu$m) $>$ 3, and $\nu$f$_{\nu}$(24 $\mu$m)/$\nu$f$_{\nu}$(0.7 $\mu$m) $>$ 10.  The f$_{\nu}$(8 \ums) are measured from the published spectra for those sources listed has having $\tau$ $>$ 0.7 and EW(6.2 \um) $<$ 0.1, or no detection of 6.2 \um.  

15. 8 absorbed AGN from the 16 sources in \citet{mar08} not duplicated in any of the samples above, selected as "obscured quasars" in the FLS field based on infrared colors and radio luminosities.  Absorption $\tau$ and the f$_{\nu}$(8 \ums) were measured from published spectra for sources without detected PAH. 

16. 5 silicate emission AGN (Table 2) from \citet{hao05}, selected as luminous, optically-classified type 1 AGN. 

17. 13 silicate emission AGN (Table 2) from \citet{sch07}, selected as known, optically-classified type 1 AGN. 

18. The silicate emission AGN PG2112+059 (Table 2) from \citet{mk07}.

19. 3 silicate emission AGN with new spectral extractions described in section 2.3, below.  Sources derive from sources with IRS spectra in the FLS and SWIRE Lockman fields having f$_{\nu}$(24 $\mu$m) $>$ 5 mJy and $R$ $>$ 20 mag. which have optical redshifts and classifications cited as type 1 AGN in the National Extragalactic Database (NED).  

20. 26 silicate emission AGN with new spectral extractions described in section 2.3.2, below. Sources are those which have optical redshifts and classifications as type 1 AGN, either as listed in NED or as determined from examination of spectra from the Sloan Digital Sky Survey (SDSS, Gunn et al. 1998), and have IRS spectra available within $Spitzer$ Legacy Program 40539 (G. Helou, P.I.), a flux limited sample of sources with f$_{\nu}$(24 $\mu$m) $>$ 5 mJy. (One source which also has $R$ $>$ 20 mag. is included above in reference 19.)

Co-ordinates, redshifts, measured values of f$_{\nu}$(8 \ums), and the resulting luminosities $\nu$L$_{\nu}$(8 \ums) for these 175 AGN are given in Tables 1 and 2.  Luminosities are determined for H$_0$ = 71 \kmsMpc, $\Omega_{M}$=0.27, and $\Omega_{\Lambda}$=0.73.

\subsection {New Observations and New Spectra} 

\subsubsection {New Obscured AGN}

The IRAS ULIRGs give large samples of absorbed AGN with z $<$ 0.3, and the sources discovered by $Spitzer$ with f$_{\nu}$(24 $\mu$m) $\ga$ 1 mJy give large samples of absorbed AGN with z $\ga$ 1.5.   For redshifts 0.3 $<$ z $<$ 1.5, existing samples are much more limited.  We undertook to enlarge such samples by seeking new sources at these intermediate redshifts.  This was done by using intermediate flux and magnitude criteria to select sources.  Because the absorbed sources have heavy extinction in the optical, and because we also wanted sources at intermediate redshift, we only searched for new sources with $R$ $>$ 20 mag.  Because we desire high luminosity sources, we only searched for new sources with f$_{\nu}$(24 $\mu$m) $>$ 5 mJy.

To obtain this new sample of absorbed AGN, we first examined all sources in the FLS \citep{fad06} and in the SWIRE Lockman Hole survey \citep{lon,lon04} to locate sources with 
f$_{\nu}$(24 $\mu$m) $>$ 5 mJy and $R$ $>$ 20 mag.  Within the two surveys, 51 sources satisfy these criteria.  Of these 51, IRS spectra are previously published (2) or available in the $Spitzer$ archive (31) for 33 sources.  We obtained new observations for an additional 14 of these sources.  We made new spectral extractions for the 14 new spectra and for the 31 unpublished spectra in the archive, giving a total of 47 IRS spectra out of the 51 sources in the complete sample.  

Most spectra have observations in the IRS Short Low module in
orders 1 and 2 (SL1 and SL2) and with the Long Low module in orders 1 and 2 (LL1 and
LL2), described in \citet{hou04}\footnote{The IRS was a collaborative venture between Cornell
University and Ball Aerospace Corporation funded by NASA through the Jet Propulsion Laboratory and the Ames Research Center}.  These give low resolution spectral
coverage from $\sim$5\,\um to $\sim$35\, \um.  Spectral extractions were done using the SMART analysis package \citep{hig04} with either v15 or v18 of the $Spitzer$ basic calibrated data products.  Extractions used all off-order observations as source background (i.e. all LL2 observations are coadded to produce the background for each nod of LL1), or alternative nod positions provide background subtraction for SL1 when SL2 was not available.  To increase S/N, spectral extractions used an extraction window of average width (perpendicular to dispersion) of 4 pixels, compared to the standard extraction width of 8 pixels that applies for flux calibrations.  These "narrow" extractions are corrected to accurate fluxes using extractions in the same fashion of the standard calibrating source Markarian 231. 

The resulting 45 spectra were examined to find those which satisfy the criteria of $\tau$ $>$ 0.7 and EW(6.2 \um) $<$ 0.1.  Of the 45, 9 sources satisfy these criteria; these are illustrated in Figure 2.  Sources 73 and 111 are from archival program 30447 (G. Fazio, P.I.), source 90 from archival program 20629 (L. Yan, P.I.), source 16 from archival program 40539 (G. Helou, P.I.), and sources 13,20,21,23, and 70 from our own new observations, program 50031.

\subsubsection {New Unobscured, Silicate Emission AGN}

IRS spectra of faint type 1 AGN having only broad and weak silicate emission features cannot determine accurate redshifts.  For this reason, there are no samples of silicate emission AGN with redshift determinations only from IRS spectra, in contrast to the large numbers of absorbed AGN which are discovered with $Spitzer$ and have redshifts from the IRS.  To be confident of the redshift for an unobscured, silicate emission AGN, we require a redshift determined from an optical spectrum.  

In order to increase the sample of silicate emission AGN, we have extracted new IRS spectra of sources which already have optical redshifts and classifications as type 1 AGN, as cited in NED.  These new sources arise from two criteria: 

\noindent 1. The sample of 45 IRS spectra described above in section 2.3.1 from the FLS and SWIRE Lockman Field with f$_{\nu}$(24 $\mu$m) $>$ 5 mJy and $R$ $>$ 20 mag. Three of these sources have optical redshifts and classifications as type 1 AGN and show silicate emission features. (All 3 also have weak PAH, EW(6.2 \ums) $<$ 0.1, to meet our criterion of negligible starburst luminosity).  These sources are listed in Table 2 as reference 19.  
 
\noindent 2. All sources from $Spitzer$ program 40539 (G. Helou, P.I.) which have optical redshifts and classifications as type 1 AGN, either as listed in NED or as determined from our examination of spectra available from the SDSS.  This $Spitzer$ Legacy Program is a flux limited sample of 330 sources with f$_{\nu}$(24 $\mu$m) $>$ 5 mJy having complete IRS low resolution spectra. We identify 25 sources which have SDSS spectra with a classification as type 1.  There are an additional 13 sources with NED citations to classifications as type 1 AGN not derived from SDSS spectra. We extracted all 38 of these spectra in order to determine those sources which show silicate emission. Of the 25 SDSS type 1 sources, 17 show silicate emission features [all also have EW(6.2 \um) $<$ 0.1].  One of these, source 16 in Table 2, satisfies the criteria of set 1, above, and is listed under reference 19; the remaining 16 SDSS unobscured AGN are listed in Table 2 as reference 20.  Of the 13 additional type 1 sources, 10 show silicate emission features [all also have EW(6.2 \ums) $<$ 0.1], and these 10 are also listed in Table 2 under reference 20, with citations to the original redshift sources.   

Our only use of the spectra for these silicate emission sources with optical redshifts is to measure the rest frame $\nu$L$_{\nu}$(8 \ums). This measurement is listed in Table 2, so these new silicate emission spectra are not illustrated.  

\section{Discussion}

\subsection{Luminosity Evolution of Most Luminous Dusty AGN}

We first present a description of the redshift dependence on luminosity for all of the AGN, obscured and unobscured, using the data in Tables 1 and 2.  This analysis is analogous to our similar analysis of the most luminous starbursts \citep{wee08}.  In both cases, the measure of luminosity is a consistent spectral feature (7.7 \um PAH for starbursts or 8 \um continuum for AGN) which can be traced over all redshifts 0 $<$ z $\la$ 2.8 using observations with the IRS.  Using a single spectral feature allows the straightforward comparison of sources over this full redshift range. 

\subsubsection{Luminosity Function of Obscured AGN}
For the obscured AGN characterized by silicate absorption, the distribution of dust continuum luminosities with redshift is shown in Figure 3.  Because the IRS spectra allow redshift determinations for AGN at high redshift having strong silicate absorption, and because many IRAS ULIRGs at low redshift are obscured AGN, there is a good distribution of these obscured AGN over all redshifts 0 $<$ z $\la$ 2.8 (triangles and crosses in Figure 3).

We quantitatively determine the form of evolution for the most luminous obscured AGN within this sample.  The form of this luminosity evolution as a function of redshift is obtained by using the most luminous source in each interval of 0.02 in log(1+z).  There are 29 such intervals in the redshift range 0 $<$ z $<$ 2.8, and all intervals except two have sources within the interval.  Three very luminous ULIRGs at low redshifts in Figure 3 (the sources whose spectra are shown in Figure 5) are excluded from the fit, as discussed further below in section 3.2.  

The linear (first order) least squares fit to the most luminous obscured AGN is given by
\begin{equation}
$$log[$\nu$L$_{\nu}$ (8 $\mu$m)] = 44.87($\pm$0.09) + 2.60($\pm$0.27) log(1+z) for $\nu$L$_{\nu}$ in ergs s$^{-1}$.$$
\end{equation}
\noindent The solid line in Figure 3 illustrates this fit to z = 2.5 for maximum luminosity within 0 $<$ z $<$ 2.8 (uncertainties in this equation are the one sigma uncertainties arising from the least squares fit).

\subsubsection{Luminosity Evolution of Unobscured, Type 1 AGN}

The sample of unobscured, silicate emission, type 1 AGN discovered with IRS observations of $Spitzer$ sources is much smaller than the sample of obscured AGN because the IRS cannot obtain an unambiguous redshift for sources with broad or weak silicate emission features. 
The unobscured, silicate emission, type 1 AGN from Table 2 are plotted in Figure 4, but we do not attempt a fit to their luminosity evolution because the redshift intervals are so sparsely sampled.  The distribution of points indicates, however, that their form of evolution is similar to that for the obscured AGN in Figure 3.  Trends are offset because of the systematically higher luminosities of the unobscured AGN, discussed further below in section 3.5.  

The distribution of the silicate emission AGN in Figure 3 matches the form of luminosity evolution for type 1 AGN in the Bootes field initially discovered as $Spitzer$ 24 \um sources and subsequently found to have optical redshifts and type 1 classifications \citep{bro07}. Using a sample of 183 sources with f$_{\nu}$(24  \ums) $>$ 1 mJy and optical redshifts 1 $<$ z $<$ 5, \citet{bro07} determine that pure luminosity evolution fits the results with evolution of the form log L(z) = log L(z=0) + 1.15z - 0.34z$^{2}$ + 0.03z$^{3}$.  This evolution curve is shown in Figure 3 as a dashed curve, normalized to the brightest unobscured AGN which has been observed in Bootes with the IRS.  The similarity in shape of this curve to those which we derived for the obscured AGN can be seen.  The \citet{bro07} function maximizes at z = 2.5. Brown et al. emphasize that this evolution agrees with that found for type 1 quasars discovered optically, as in \citet{ric06}.  

The demonstrated similarity in evolution of both obscured and unobscured AGN is one of our most important results.  It is one source of evidence, with more discussed below, that there are no fundamental, intrinsic differences among the most luminous AGN despite the very differing spectral signatures which divide obscured and unobscured AGN.

\subsubsection{Comparison to Evolution of Starbursts}

Another important result of the analysis above is that the evolution factor of (1+z)$^{2.6}$ for obscured AGN agrees with that determined for the most luminous starbursts \citep{wee09}.  The form of evolution is determined the same way in both cases, taking the most luminuous source in each interval 0.02 of log (1+z) and extending to z = 2.6.  In both cases, the most luminous sources were located within a variety of $Spitzer$ spectroscopic programs, and a single spectral diagnostic was used to trace luminosities over all redshifts from 0 to 2.6. 

For starbursts, the parameter which was used in \citet{wee09} to measure evolution is the luminosity $\nu$L$_{\nu}$ (7.7 $\mu$m) for the 7.7 \um PAH emission feature associated with starbursts. This luminosity relates to the star formation rate (SFR) for starbursts, so luminosity evolution also represents evolution in the maximum SFR, which was found to scale as log(SFR) = 2.1($\pm$0.3) + 2.5($\pm$0.3) log(1+z), for SFR in \mdot.  

We note also that the form of evolution we derive for obscured and unobscured AGN and for starbursts is the same within the uncertainties as found for optically-luminous galaxies. The comprehensive summary by \citet{hop04} of evolution parameters derived from optical and radio constraints shows evolution going as (1+z)$^{2.9}$ and being almost entirely luminosity evolution, with $\pm$0.5 uncertainty in the exponent. 

The sources which we use to track AGN evolution and starburst evolution are from completely different samples, each sample defined by the presence (starburst) or absence (AGN) of the PAH emission feature in the IRS spectra.  Our observational result is that these two categories of luminous, dusty galaxies have evolved in the same way.  The resulting similarity in evolution implies a close relationship between luminous AGN and luminous starbursts.  What is the nature of this relationship? 

Attempts to understand the local ULIRGs resulted in the suggestion that the starburst and AGN phenomena in luminous, dusty galaxies occur in the same galaxy but at different times \citep{san88}.  In a simple summary of this scenario, a starburst is triggered by the inflow of a new supply of molecular gas caused by a galactic interaction or merger.  Some of the gas reaches the central massive black hole and causes an increase in its luminosity, but the AGN is not seen until the radiation from the AGN dissipates the surrounding gas and dust, which may also quench the starburst.  A similar scenario has been suggested for the extremely luminous, dust obscured galaxies at high redshift, including those in the present paper \citep{dey08}.  In this case, contrasted to the local ULIRGs, the molecular material is collecting about a primordial galaxy in the process of formation.  

That the luminous AGN and starburst populations track each other so closely in luminosity evolution does not require, however, that either phenomena causes the other.  If both luminous AGN and luminous starbursts arise because of galactic interactions or other events which supply fresh gas and dust into the gravitational potential of a galaxy, both can be triggered by the same event.  This triggering does not require that the AGN affects the starburst, or that the starburst affects the AGN; they can be coincidental in the same galaxy because a galactic interaction stimulates both.  Whatever the process, these $Spitzer$ IRS results show that energizing sources for dusty ULIRGS, whether starburst or AGN, have declined together through the past 11 billion years, as the typical luminosity has faded by a factor of $\sim$ 20.  

\subsection{The Extreme Local ULIRGs}

One exception should be noted to the similarities described above among the evolution of both categories of AGN and of starbursts.  For starbursts and unobscured (type 1) AGN, there are no sources with z $<$ 0.5 having luminosities comparable to the most luminous sources with z $>$ 2.  There is a distinctive difference for the most luminous, obscured AGN.  The three most luminous, obscured ULIRGs in Figure 3, all with z $<$ 0.6, are systematically more luminous by a factor of $\sim$ 3 than other local ULIRGs and come within a factor of 3 of the most luminous absorbed AGN found anywhere in the Universe.  

The rest-frame spectra of these three sources (all discovered by IRAS) are shown in Figure 5, labeled by the running numbers in Table 2.  The spectra illustrate that all of these are deeply absorbed, so that the high luminosities $\nu$L$_{\nu}$(8 $\mu$m) arise despite extinction that may decrease the apparent 8 $\mu$m luminosities.  The weakness of the 6.2 \um PAH feature is evidence that the high luminosities at 8 \um do not arise from a starburst component that enhances the apparent 8 $\mu$m luminosity with 7.7 \um PAH emission.  The spectra in Figure 5 also show how silicate absorption $\ga$ 8 $\mu$m and absorption by ices $\la$ 8 $\mu$m produces a localized continuum peak at $\sim$ 8 $\mu$m.  The average shape of these spectra shown in Figure 1 is very similar to the spectra of the most luminous absorbed AGN at z $\sim$ 2 \citep[e.g. ][]{hou05,pol08,saj08}.

For luminous starbursts and unobscured AGN, luminosity evolution can be explained by a systematic change in the amount of gas that falls deep within the potential well of the galaxy; unobscured AGN luminosity is controlled by the fraction of this gas that approaches the central massive black hole, and starburst luminosity is controlled by the total amount of gas in the circumnuclear molecular clouds.  As a consequence, the luminosity evolution of type 1 AGN and of starbursts both is the result of having decreasing amounts of accreting material at lower redshifts.  

For the luminous, obscured AGN, the presence of three low redshift sources which depart from the systematic trend indicates that there may be an additional factor other than the amount of accreted material which sometimes controls the luminosity $\nu$L$_{\nu}$(8 $\mu$m) of these sources.  It is notable, however, that the unusually high luminosities of these 3 sources when measured by $\nu$L$_{\nu}$(8 $\mu$m) are not so unusual when measured by total infrared luminosity L$_{IR}$.  The 2 IRAS ULIRGs in Figure 5 have measures of L$_{IR}$ (sources 3 and 32, excluding the FSC source), and the ratio L$_{IR}$/$\nu$L$_{\nu}$(8 $\mu$m) is shown in Figure 6.  The 2 unusually luminous ULIRGs have distinctly small values for this ratio, indicating that they are more similar to other ULIRGs when measured in L$_{IR}$. 

One possible explanation for the unusually large $\nu$L$_{\nu}$(8 $\mu$m) and unusually small ratio L$_{IR}$/$\nu$L$_{\nu}$(8 $\mu$m) may be the location and distribution of the absorbing dust clouds.   Absorbing clouds concentrated unusually close to the AGN could give rise to warmer dust and preferential emission at shorter wavelengths, compared to sources with cooler, more distant clouds.  We can consistently interpret our general results for both obscured and unobscured AGN in context of the simple "unified model" (section 3.5, below) containing a dusty, absorbing torus with comparable opening angles in all AGN.  These three sources indicate, however, that this model becomes more complex for the most heavily absorbed sources \citep[e.g. ][]{lev07,ima07,pol08}.  For this reason, we determine the form of luminosity evolution in section 3.1.1. for obscured sources without including these three anomalous sources.

\subsection{Total Infrared Luminosities of AGN and Starbursts}

We have utilized $\nu$L$_{\nu}$(8 $\mu$m) as a measure of continuum luminosity from emitting dust because of observational necessity; this parameter is easily measured over a wide range of redshifts.  Of course, a more fundamental measurement is the total dust-emitted luminosity, L$_{IR}$, which is dominated by luminosity arising at far-infrared wavelengths from dust that is cooler than the dust which produces the mid-infrared luminosity at 8  \ums.   For those bright sources which have total infrared observations with IRAS, we can determine the ratio L$_{IR}$/$\nu$L$_{\nu}$(8 $\mu$m).  

For the obscured ULIRGs in Table 1, the result is shown in Figure 6, using the $\nu$L$_{\nu}$(8 $\mu$m) from Table 1 and the L$_{IR}$ from \citet{far07} or \citet{ima07}. The median ratio log [L$_{IR}$/$\nu$L$_{\nu}$(8 $\mu$m)] = 0.95$\pm$0.25, which gives that the median correction for obscured ULIRGs is log L$_{IR}$ = log $\nu$L$_{\nu}$(8 $\mu$m) + 0.95$\pm$0.25.  (Because of the small number of points, we do not attempt to fit a relation having luminosity dependence.).  Assuming this also applies to other obscured AGN in our sample, this yields that for obscured AGN,
\begin{equation}
$$log L$_{IR}$(AGN$_{obscured}$) = log $\nu$L$_{\nu}$(8 $\mu$m) - 32.63$\pm$0.25, for L$_{IR}$ in \ldot~and $\nu$L$_{\nu}$(8 $\mu$m) in ergs s$^{-1}$.$$
\end{equation}

This is simply an empirical result based on local IRAS ULIRGs having observed data that include far-infrared wavelengths.  Efforts have begun to obtain total luminosities for the high redshift, absorbed sources discovered by $Spitzer$ (including many sources in Table 1) using submillimeter observations \citep{saj08}; these authors give results in the rest frame compared to the broad band measurement that would be made with the 8 \um filter of the $Spitzer$ Infrared Array Camera (IRAC).  The empirical result in \citet{saj08} is log L$_{IR}$ = (2.79$\pm$ 0.36) + (0.83$\pm$ 0.03)log $\nu$L$_{\nu}$(IRAC 8  \ums), for luminosities in \ldot.  

Applying the synthetic photometry tool in SMART to the average spectrum of the absorbed source in Figure 1 indicates that the flux density observed with the IRAC 8 \um filter is 0.68 of the monochromatic f$_{\nu}$(8 $\mu$m) for an absorbed source with a spectrum as in Figure 1. Using this transformation, the result from Sajina et al. becomes log L$_{IR}$ = 2.65  + 0.83 log $\nu$L$_{\nu}$(8 $\mu$m).  The median log $\nu$L$_{\nu}$(8 $\mu$m) in Figure 6 is 45.0 (ergs s$^{-1}$) or 11.41 (\ldot), which yields log L$_{IR}$ = 12.12 using the Sajina et al. transformation. Using our transformation in Equation 3, a source with log $\nu$L$_{\nu}$(8 $\mu$m) = 45.0 (ergs s$^{-1}$) would have 
log L$_{IR}$ = 12.37 (\ldot).  The difference of 0.25 between these two independent determinations of log L$_{IR}$ is well within the stated uncertainties of the two relations.  This comparison gives confidence that the transformation between L$_{IR}$ and $\nu$L$_{\nu}$(8 $\mu$m) is similar among all of the absorbed AGN in Table 1 and Figure 3.
  
For the unobscured, silicate emission, type 1 AGN in Table 2, we determine L$_{IR}$ for those which have total infrared luminosities measured at all IRAS wavelengths, taken from the study of PG quasars in \citet{sch07} and including 3C 273 from \citet{hao05} and \citet{far07}. Source luminosities L$_{IR}$ are in Table 2 and the comparison to $\nu$L$_{\nu}$(8 $\mu$m) are in Figure 7.  Asterisks show values for sources without measured PAH features and diamonds for sources with weak PAH features, although there is no systematic difference in the correction.  The median ratio log [L$_{IR}$/$\nu$L$_{\nu}$(8 $\mu$m)] = 0.74$\pm$0.15.  This indicates that, for unobscured AGN, 
\begin{equation}
$$log L$_{IR}$(AGN1) = log $\nu$L$_{\nu}$(8 $\mu$m) - 32.84$\pm$0.15 for L$_{IR}$ in \ldot~and $\nu$L$_{\nu}$(8 $\mu$m) in ergs s$^{-1}$.$$
\end{equation}

For starbursts calibrated in \citet{hou07}, the analogous result is log L$_{IR}$ = log $\nu$L$_{\nu}$(7.7 $\mu$m) + 0.78$\pm$0.2 in ergs s$^{-1}$, which gives that
\begin{equation}
$$log L$_{IR}$(SB)= log $\nu$L$_{\nu}$(7.7 $\mu$m) - 32.80$\pm$0.2 for L$_{IR}$ in \ldot~and $\nu$L$_{\nu}$(7.7 $\mu$m) in ergs s$^{-1}$.$$
\end{equation}

\noindent The corrections to L$_{IR}$ appear similar for AGN and starbursts, but this is not the case if only the dust emission is measured.  For starbursts, the $\nu$L$_{\nu}$(7.7 $\mu$m) measures PAH emission.  The actual dust continuum is much weaker.  For a pure starburst, such as the lower spectrum in Figure 1, the dust continuum can be estimated by fitting the continuum on either side of the broad PAH complex, and this continuum is about a factor of ten weaker than the peak at 7.7 \um \citep[e.g.][]{bra06}.  A more appropriate rest wavelength at which to compare intrinsic dust continua is 15  \ums, beyond the silicate absorption and not contaminated by PAH.  But we cannot observe this rest wavelength with the IRS for sources with z $\ga$ 1.2. 
 
Using the transformations in equations (2), (3), and (4) together with the form of evolution in equation (1) from section 3.1.1., we can determine the total luminosity L$_{IR}$ as a function of redshift for the most luminous AGN.  This can be compared with the result previously derived in the same fashion for the most luminous starbursts \citep{wee08}. Results are: 

\noindent a) For the most luminous obscured AGN, 
\begin{equation}
$$log L$_{IR}$(AGN$_{obscured}$) = 12.3$\pm$0.25 + 2.6($\pm$0.3)log(1+z) for L$_{IR}$ in \ldot.$$ 
\end{equation} 
\noindent b) For the most luminous unobscured, type 1 AGN, 
\begin{equation}
$$log L$_{IR}$(AGN1) = 12.6$\pm$0.15 + 2.6($\pm$0.3)log(1+z) in \ldot.$$ 
\end{equation} 
\noindent c) For the most luminous starbursts,
\begin{equation}
$$log L$_{IR}$(SB) = 11.8$\pm$0.3 + 2.5($\pm$0.3)log(1+z) in \ldot.$$ 
\end{equation} 

The results in equations (5) and (6) summarize once again the fundamental similarities between obscured and unobscured AGN, with their bolometric luminosities being the same within the uncertainties.  The results also indicate that the most luminous sources in the Universe arise from AGN and not from starbursts.   
\subsection{Luminosity Functions for the Most Luminous Dusty Galaxies at z = 2.5}
\subsubsection{Luminosity Function of Obscured AGN}
Since the discoveries of IRAS, it has been known that the most luminous galaxies have their luminosity dominated by continuum radiation in the infrared, and this led to the definition of ULIRGs \citep {soi86,san88}.  The sources summarized in this paper represent the most luminous AGN known at any redshift, but our data do not extend sufficiently deep to determine luminosity functions extending through a large range of luminosity. 

The present sample can only determine the brightest end of the luminosity function at high redshift for the obscured AGN.  The obscured AGN which have been found in the Bootes, FLS, and Lockman Hole survey areas represent a reasonably complete sample at z $\sim$ 2 for sources with f$_{\nu}$(24 $\mu$m) $\ga$ 1 mJy within a total sky area of 22 deg$^{2}$.  This is because of the many programs described in section 2.2 to discover these sources.  The most significant selection effect depends on redshift. The peak of the spectrum at 8 \um (Figure 1) means that the f$_{\nu}$(24 $\mu$m) is dependent on redshift as this peak moves through the MIPS 24 \um filter band.  This explains why so many sources have been found with z $\sim$ 2. 

In Figure 8, the distribution of redshifts and source luminosities for the obscured AGN in Table 1 is shown for 1.5 $<$ z $<$ 3.0.  This Figure demonstrates the preferential selection of sources with 1.7 $<$ z $<$ 2.7.  We use this distribution of sources to determine a luminosity function applying to the luminosities which are shown. 

We divide the results in Figure 8 into two redshift windows, 1.7 $<$ z $<$ 2.2 and 2.3 $<$ z $<$ 2.7, to make independent estimates of the luminosity function in each window.  Sources over the luminosity range 12.9 $<$ log L$_{IR}$ $<$ 13.9 (\ldot) are counted in each interval of 0.4 in log L$_{IR}$ for each redshift window.  Luminosities for sources in the lower redshift window centered at z = 1.95 are scaled by an evolution factor (1+z)$^{2.5}$ to match luminosities in the higher redshift window centered at z = 2.5, so that the resulting luminosity function applies for z = 2.5.  Total volumes contained within the two redshift intervals are determined by scaling to the 22 deg$^{2}$ included in the surveys which discovered these sources and using the cosmology described in the notes to Table 1. 

The resulting luminosity function is shown in Figure 9.  For comparison, we also show the bright end of the local luminosity function for all IRAS galaxies from Soifer, Neugebauer and Houck (1987), scaled to z = 2.5 using the evolution factor (1+z)$^{2.6}$ which we have determined for the dusty AGN.  The IRAS luminosity function includes all galaxies and makes no distinction among obscured AGN, unobscured AGN, or starbursts.  From the $Spitzer$ results for ULIRG spectra, in section 3.2, we can conclude that the IRAS sources which define the bright end are obscured AGN.  The agreement between the extrapolated IRAS luminosity function and our new determination from $Spitzer$ sources is independent confirmation of the validity of the evolution factor, and of the validity of the empirical transformation from $\nu$L$_{\nu}$(8 $\mu$m) to L$_{IR}$. 

\subsubsection{Comparison to Unobscured AGN and Starbursts}

To understand the total infrared luminosity density contributed by obscured AGN, unobscured AGN, and starbursts, it is necessary to understand the relative luminosity functions and space densities for these three categories of sources.  Results on this question will be further refined as increased understanding develops on the classification and redshift of the large numbers of sources in $Spitzer$ surveys, and how these sources relate to AGN and starbursts found with other techniques.

For now, we can qualitatively scale the luminosity function in Figure 9 to estimate the analogous luminosity functions for unobscured AGN and starbursts just by comparing the fluxes and numbers of such sources found in the same surveys that reveal the obscured AGN.  Among the programs for IRS follow up of Spitzer sources with f$_{\nu}$(24 $\mu$m) $\ga$ 1 mJy chosen without considering the near infrared spectral energy distributions \citep[e.g.][]{hou05, yan07, wee06a,wee09}, the numbers of obscured AGN (found with silicate absorption) and the number of unobscured AGN (featureless or silicate emission sources) are comparable.  The total number of these AGN near z $\sim$ 2 is approximately the same as the total number of starbursts classified by PAH emission.  PAH sources dominate, however, in samples deliberately selected using near infrared IRAC colors to seek the redshifted 1.6 \um stellar absorption feature \citep{wee06c,far08}. 

More quantitative results comparing the numbers of AGN and starbursts arise by classifying the spectral shapes of sources having MIPS and IRAC fluxes.  Sources with power law spectral energy distributions consistently have other characteristics of AGN \citep{pol08,brn06,don08}.  The AGN fraction determined in this way decreases from $\sim$ 40\% at f$_{\nu}$(24 $\mu$m) $>$ 1 mJy to $\sim$ 10\% at f$_{\nu}$(24 $\mu$m) $\sim$ 0.1 mJy \citep{don08}.  Such surveys also indicate that the numbers of obscured and unobscured AGN are similar, to the survey limits \citep{hic08}. 

We can conclude, therefore, that the space densities at z $\sim$ 2 for the most luminous obscured AGN, unobscured AGN, and starbursts are similar.  Their relative luminosity densities then scale as the relative luminosities.  These relative L$_{IR}$ can be estimated by scaling the brightest examples at z = 2.5, as in Figure 1, and using the transformations to L$_{IR}$ in equations 2-4.  Determining from Figure 1 that f$_{\nu}$(8 $\mu$m,AGN1) : f$_{\nu}$(8 $\mu$m,AGN$_{obscured}$) : f$_{\nu}$(7.7 $\mu$m,SB) = 6 : 2.5 : 1, the result is L$_{IR}$(AGN1) : L$_{IR}$(AGN$_{obscured}$) : L$_{IR}$(SB) = 5.5 : 3.7 : 1.0.  

If relative luminosity densities scale the same as these relative luminosities, this result means that AGN of both categories together produce about ten times the luminosity density as do starbursts.  We emphasize, however, that this conclusion applies only at the very highest luminosities and does not represent a measurement over the full luminosity functions of AGN and starbursts.

\subsection{Unified Model and Extinction of AGN}

The transformations from $\nu$L$_{\nu}$(8 $\mu$m) to L$_{IR}$ used above are empirical and do not apply any corrections for extinction of $\nu$L$_{\nu}$(8 $\mu$m).  Nevertheless, understanding extinction at 8 \um is important for understanding the geometry of the sources and the amount of dust they contain.  Extinction can be estimated under the simple assumption of an absorbing screen, using measurements of absorption $\tau$ at 9.7 \um and scaling to absorption by the same dust at 8 \um \citep{dra01}.  There are two primary uncertainties with this approach.  One is that absorption is not taking place in a screen, but is taking place because of complex radiative transfer effects within dust clouds which are hotter on one side and cooler on the other \citep[e.g.][]{lev07,ima07}.  The other uncertainty arises because the 9.7 \um absorption is often so deep that an accurate measure of $\tau$ is not possible.

A number of studies have verified the validity of the "unified model" of AGN for the infrared dust emission and absorption \citep[e.g.][]{shi07,hao05,hic08,mai07}. Originally conceived to explain the differences between optical classification of type 1 AGN and type 2 AGN as arising only from orientation effects \citep{ant85}, the model can be applied to the infrared dust continuum.  In this case, the observed $\nu$L$_{\nu}$(8 $\mu$m) depends on viewing angle.  If absorbing clouds are viewed from the cool side, then L$_{IR}$/$\nu$L$_{\nu}$(8 $\mu$m) is measured by the value for obscured sources.  If viewed so that the hot side of the clouds is directly observed, L$_{IR}$/$\nu$L$_{\nu}$(8 $\mu$m) is measured by the value for unobscured sources.

If both unobscured and obscured AGN have intrinsically the same luminosities L$_{IR}$, then the transformations from $\nu$L$_{\nu}$(8 $\mu$m) to L$_{IR}$ derived above in section 3.3 differ only because the $\nu$L$_{\nu}$(8 $\mu$m) suffers more extinction for the obscured sources.  The total L$_{IR}$ is independent of viewing angle because it is dominated by far infrared luminosity which is not subject to significant extinction regardless of viewing angle.   

Comparing the transformations from $\nu$L$_{\nu}$(8 $\mu$m) to L$_{IR}$ in equations (2) and (3), the results show that the correction is different by a factor of 1.6 between obscured and unobscured AGN, in the sense that $\nu$L$_{\nu}$(8 $\mu$m) is more luminous relative to L$_{IR}$ in unobscured, type 1 AGN compared to the obscured AGN.  This result is as expected in the unified model if the obscured sources are viewed through dust clouds and indicates a typical extinction at 8 $\mu$m by a factor of 1.6 in the obscured AGN, or that $\sim$ 60\% of the intrinsic 8 \um luminosity emerges from obscured sources. 

An independent estimate of the extinction at 8 \um arises by comparing the observed luminosities of obscured and unobscured AGN, as in Figures 1 and 4. 
If the intrinsic luminosities for obscured, silicate absorbed sources and the unobscured, silicate emission sources are the same, the scaling difference in $\nu$L$_{\nu}$(8 $\mu$m) is a measure of how much luminosity has been extincted at 8 \um for the absorbed sources.  This scaling difference in Figures 1 and 4 is about 2.5, i.e. the unabsorbed sources are typically two to three times as luminous at 8 \um as the absorbed sources.  This result implies that $\sim$ 40\% of the intrinsic 8 \um luminosity emerges from the obscured sources.

\subsection{Tracing Dusty AGN and Starbursts to z = 6}

The presence of large amounts of dust within AGN and starbursts in the early Universe raises many questions.  Does the luminosity evolution observed to z = 2.6 continue to higher redshifts?  When and how was the dust formed?  When was the Universe most obscured?  Were massive black holes present before or after dust formation?  Which came first, starbursts or AGN?  Progress toward answering such questions depends on tracking the dust content of the Universe as a function of redshift.  

The determination of luminosity evolution discussed in previous sections extends to z = 2.6 only because of the limit in the spectral response of the IRS.  At z = 2.6, the 8 \um peak of the absorbed continuum is at $\sim$ 29 \um, and the deepest absorption is at $\sim$ 35 \um.  For faint sources with poor S/N, these are approximately the reddest wavelengths for which absorbed sources can be confidently identified spectroscopically.  A crucial question, therefore, is whether the luminosity evolution observed to z = 2.6 continues beyond that redshift.  Even though infrared spectra for redshift measurements cannot be obtained, we can ask whether such sources would exist within the $Spitzer$ MIPS surveys at 24 \um.  This is done by using the results described above regarding the luminosities and evolution of these sources to predict the fluxes which would be observed at 24 \um as redshift increases. 

The results are shown in Figure 10. Asterisks are the most luminous unobscured, type 1 AGN, crosses are the most luminous absorbed AGN, and diamonds are the most luminous starbursts, all scaled to the brightest known examples shown in Figure 1.  For each set, the upper curve shows expected f$_{\nu}$(24 $\mu$m) if source luminosities continue to scale as (1+z)$^{2.5}$ to z = 6, and the lower curve shows expected f$_{\nu}$(24 $\mu$m) if there is no luminosity evolution for z $>$ 2.5. (The latter is the result for type 1 AGN by \citet{bro07}, who determine that the space density of luminous type 1 sources maximizes at z = 2.5.)  The horizontal line in Figure 9 is the MIPS limit of f$_{\nu}$(24 $\mu$m) = 0.3 mJy for typical wide area surveys such as in Bootes, the FLS, and SWIRE. (Smaller area surveys have reached f$_{\nu}$(24 $\mu$m) $<$ 0.1 mJy.) 

Results in Figure 10 show that even if luminosity evolution continues for z $>$ 2.5, we could not expect these MIPS surveys to identify starbursts with z $>$ 3.  The rapid fading of starbursts arises in part because the strong PAH 7.7 \um emission feature moves longward of the MIPS bandpass.  $Spitzer$ studies are revealing large numbers of dusty starbursts with z $<$ 3 by various survey techniques and IRS follow up, and these studies should eventually lead to a well quantified measure of star formation densities for z $<$ 3.  For now, tracking dust to earlier epochs must rely on AGN. 

Figure 10 shows that both obscured AGN and unobscured (type 1) AGN should be found with f$_{\nu}$(24 $\mu$m) $>$ 0.3 mJy to z $\sim$ 6, even without continued luminosity evolution for z $>$ 2.5.  In fact, a number of type 1 AGN have already been discovered using f$_{\nu}$(24 $\mu$m) in the Bootes field at these redshifts \citep{bro07,coo06}.  These are shown as squares in the Figure.  (Redshifts are not published for the quasars in Brown et al., but redshifts and fluxes for sources are shown in Figure 8 of that paper, and all sources with z $>$ 3 are shown in Figure 9.)  That the f$_{\nu}$(24 $\mu$m) for all of these unobscured AGN in Figure 10 fall below the extrapolations from the brightest source at z = 2.5 is further evidence that the luminosity evolution of unobscured AGN turns down for z $>$ 2.5, as already concluded from optical data.

It is important to emphasize that the evidence for this turnover of luminosity evolution in Figure 10 derives from the luminosity of the dust-emitting continuum.  One possible suggestion for the maximal optimal luminosities of type 1 AGN at z $\sim$ 2.5 when observed in the rest-frame ultraviolet is that dust and extinction were less at that epoch.  But the evidence of the present analysis is that dust luminosity for unobscured AGN also shows a similar maximum, so the explanation for this maximum is not the absence of dust.  

It is crucial to find more high redshift examples of either category of AGN, because these trace dust to the most distant epochs at which it can be detected.  The presence of the two quasars in Figure 9 from \citet{coo06} with z = 5.5 is evidence that dust existed when the Universe was less than 1 billion years old. 

It will be much more difficult to find obscured AGN for z $>$ 3 than to find the unobscured, type 1 AGN.  The obscured AGN are much fainter optically than the unobscured AGN; based on the $R$ magnitudes of the obscured AGN at z $\sim$ 2.5, we would expect $R$ $\ga$ 25.  Obscured AGN with z $>$ 3 may explain many of the sources having f$_{\nu}$(24 $\mu$m) $\ga$ 1 mJy and $R$ $\ga$ 25 mag which show no features in the IRS spectra \citep{wee06b}.  The best candidates for such sources are those objects with power law spectra in the MIPS and IRAC bands but optical magnitudes fainter than 25.  The only way to determine redshifts for these will be with optical and near-infrared spectroscopy, but this is challenging at such magnitudes \citep[e.g.][]{brn07,des08}. \

\subsection{Galaxy Formation at High Redshift}

That galaxy luminosities peak at z $\sim$ 2.5 indicates that this epoch corresponds to a maximum for the assembly of massive galaxies \citep[e.g.][]{dey08}.  We can use the numerical values for luminosities and SFRs derived above to compare with other, independent conclusions concerning the nature of the most luminous galaxies at such redshifts.  

The evolution of maximum SFR for starbursts to z = 2.5 reviewed in section 3.3 and the absence of starbursts with greater SFR at z $>$ 2.5 discussed in section 3.6 indicate that the maximum SFR during the epoch 2.5 $<$ z $<$ 6 is SFR $\la$ 3500 \mdot.  \citet{bus09} derive stellar masses for dust obscured galaxies at z $\sim$ 2 using HST observations that resolve the extended stellar component of the galaxy.  Bussman et al. determine a median lower limit to the stellar mass of 3x10$^{10}$ M$_{\odot}$ (depending on uncertainties for extinction corrections) for 30 obscured galaxies similar to the sources in Table 1.   At the maximum SFR which we derive, such a mass could accumulate within $\sim$ 10$^{7}$ yr.  This conclusion indicates that starbursts at the luminosities we observe need have only short duration to explain the observed stellar masses of galaxies at z $\sim$ 2.5.  

We can also ask whether the necessary mass for the central massive black hole (MBH) presumed to power the accreting AGN is consistent with the SFRs, as required if the remains of stars collect within the galactic nucleus to form the accreting MBH.  Many HST observations of local galaxies indicate a consistent and systematic result that every galaxy contains a central MBH with mass $\sim$ 0.2\% of the total virial mass of the galaxy \citep[e.g.][]{fer06} (these authors suggest the term "Central Massive Object", or CMO, to allow a concentration of smaller black holes rather than a single MBH). 

Equations (5) and (6) indicate that the most luminous AGN at z = 2.5 has L$_{IR}$ = 10$^{13.9}$ \ldot.  If the Eddington limit applies, 10$^{38}$ ergs s$^{-1}$ is produced per solar mass of the accreting massive black hole (MBH).  In this case, the minimum accreting mass for the most luminous AGN at z = 2.5 is then 3x10$^{9}$ M$_{\odot}$.  This number for the high redshift, dusty AGN is the same as the most massive local CMO observed by HST, which is shown in \citet{fer06} to have a CMO mass of 3x10$^{9}$ M$_{\odot}$. 

Using the 0.2\% scaling of Ferrarese et al., this result for the most luminous AGN indicates a required total virial mass of 1.5x10$^{12}$ M$_{\odot}$ for the most luminous obscured galaxies.  If we assume that 90\% of this virial mass is in the form of non-baryonic, dark matter, the required stellar mass is 1.5x10$^{11}$ M$_{\odot}$, a number only 5 times larger than the median lower limit for the observed stellar masses in \citet{bus09}.  Starbursts at the maximum SFR of 3500 \mdot could accumulate this stellar mass in 5x10$^{7}$ yr.  

Our various conclusions regarding the luminosities and SFRs of the high redshift AGN and starbursts discovered by $Spitzer$ are consistent, therefore, with the interpretation that these sources represent the most massive galaxies in the process of formation.  Results are also consistent with the interpretation that the central, compact massive objects which power AGN were in place by the epoch when these galaxies had maximum luminosity.

\section{Summary and Conclusions}

Continuum luminosities $\nu$L$_{\nu}$(8 $\mu$m) arising from dust emission are given for 115 obscured AGN and 60 unobscured (type 1) AGN with 0.005 $<$ z $<$ 3.2 using measurements with the $Spitzer$ IRS.  Obscured sources are defined by having optical depth $>$ 0.7 in the 9.7  $\mu$m silicate absorption feature and unobscured (type 1) AGN by having silicate in emission. 

These AGN include the most luminous sources known in the Universe as measured by $\nu$L$_{\nu}$(8 $\mu$m).  Luminosity evolution for the most luminous obscured AGN is found to scale as (1+z)$^{2.6}$ to z = 2.8.  For the more limited sample of unobscured, type 1 AGN, the scaling with redshift is similar, but the $\nu$L$_{\nu}$(8 $\mu$m) is approximately 3 times more luminous at all redshifts compared to the obscured AGN.  The evolution factor for AGN is the same as previously found in a similar way with IRS results for the most luminous starbursts in the Universe. 

Using both obscured and unobscured AGN having total IRAS fluxes, empirical calibrations are found between $\nu$L$_{\nu}$(8 $\mu$m) and L$_{IR}$.  Combining these calibrations with the redshift trends and using a similar result previously determined for starbursts, we conclude that total infrared luminosities for the most luminous galaxies scale with redshift as:

\noindent log L$_{IR}$(AGN$_{obscured}$) = 12.3$\pm$0.25 + 2.6($\pm$0.3)log(1+z) for L$_{IR}$ in \ldot, 

\noindent log L$_{IR}$(AGN1) = 12.6$\pm$0.15 + 2.6($\pm$0.3)log(1+z), and

\noindent log L$_{IR}$(starburst) = 11.8$\pm$0.3 + 2.5($\pm$0.3)log(1+z). 

\noindent Comparisons of obscured and unobscured AGN are consistent with both types having the same total luminosities with differences arising only from orientation effects, as expected in the unified model for AGN.  The obscured AGN contain very dusty clouds which extinct about 50\% of the intrinsic luminosity at 8 $\mu$m.  

The luminosity function of obscured AGN is determined for the highest luminosities at z = 2.5, and this luminosity function agrees with that for local IRAS ULIRGs if they are scaled by (1+z)$^{2.5 }$ luminosity evolution.  At the highest luminosities (L$_{IR}$ $\ga$ 10$^{13}$\ldot), obscured and unobscured AGN together contribute about 10 times more luminosity density to the early Universe than do the most luminous starbursts. 

For comparison with source fluxes found in wide-field $Spitzer$ MIPS surveys, extrapolations of observable f$_{\nu}$(24 $\mu$m) to z = 6 are made using the most luminous examples currently known.  Both obscured and unobscured AGN should be detected to z $\sim$ 6 with f$_{\nu}$(24 $\mu$m) $>$ 0.3 mJy, even without continued luminosity evolution for z $>$ 2.5. Type 1 AGN discovered using dust luminosity observed at 24 \um show a maximum in luminosity at z $\sim$ 2.5, similar to the redshift of maximum luminosity derived from optically-discovered quasars.  In contrast to AGN, the most luminous starbursts cannot be detected with f$_{\nu}$(24 $\mu$m) $>$ 0.3 mJy for z $>$ 3, even with continued luminosity evolution.  

The maximum luminosities for dusty AGN at z = 2.5 imply a minimum accreting mass of 3x10$^{9}$ M$_{\odot}$.  The observed stellar masses of such galaxies could accumulate in $\la$ 5x10$^{7}$ yr at the maximum star formation rates which are observed for dusty starburst galaxies.

\acknowledgments
We thank L. Sargsyan for help with spectral analysis and SDSS spectra, D. Farrah for providing a list of SWIRE Lockman sources, and D. Barry for continuing help with data management. This work is based primarily on observations made with the
Spitzer Space Telescope, which is operated by the Jet Propulsion
Laboratory, California Institute of Technology, under NASA contract
1407. Support for this work by the IRS GTO team at Cornell University was provided by NASA through Contract
Number 1257184 issued by JPL/Caltech. This research has made use of the NASA/IPAC Extragalactic Database (NED) which is operated by JPL/Caltech under contract with NASA.

\clearpage

\begin{deluxetable}{ccccccc} 
\tablecolumns{7}
\tabletypesize{\footnotesize}

\tablewidth{0pc}
\tablecaption{Continuum Luminosities for Obscured AGN}
\tablehead{
 \colhead{Source} &\colhead{Name}& \colhead{J2000 coordinates} &\colhead{z} & \colhead{f$_{\nu}$(8 $\mu$m)\tablenotemark{a}}& \colhead{log $\nu$L$_{\nu}$(8\,\um)\tablenotemark{b}}& \colhead{Ref.\tablenotemark{c}} 
}
\startdata



1& IRAS 00091-0738 & 001143.3-072207 	& 0.118	& 52 & 44.79& 2\\
 2& IRAS 00188-0856 & 002126.5-083926  & 0.128	& 53	 & 44.87 & 2\\
3&IRAS 00397-1312 &004215.5-125602 &0.262 &155 &45.99& 1\\
4& IRAS 01004-2237 & 010250.0-222157 & 0.118 & 92 & 45.04& 2\\
5& IRAS 01166-0844 & 011907.7-082910 	& 0.118	& 23& 44.44& 2\\ 
6& IRAS 01298-0744 & 013221.4-072908 	& 0.136	& 83 & 45.12& 2\\ 
7 &IRAS 06035-7102 &060254.0-710310 &0.079 &132 &44.84&1\\
8 &IRAS 06206-6315 &062101.2-631723 &0.092 &56&44.60&1\\
9& IRAS 08572+3915 & 090025.4+390354 & 0.058 & 560 & 45.20 & 2\\
10 &UGC 5101 &093551.7+612111.3 &0.039 &278&44.54&1\\
11 & FSC 09425+1751 & 094521.4+173753	& 0.130	& 47.9& 44.85& 3\\
12& NGC3281	& 103152.1-345113	& 	0.0107	& 1900 & 44.26& 4\\
13& SWIRE& 103303.37+572050.8			& 0.50  &  2.5 & 44.78& 6\\
14& IRAS 10378+1108 & 104029.2+105318	& 0.136	& 28 & 44.65& 2\\ 
15& SWIRE & 104314.93+585606.3	&2.25   &1.3  &45.78&12\\				
16& SWIRE & 104325.62+581852.3		& 0.67   & 3.5	& 45.19& 6\\
17&  SWIRE & 104407.97+584437.0	& 0.56   & 3.5 & 45.03 &12\\			
18& SWIRE & 104528.29+591326.7		& 2.31   & 3.4	 & 46.21 & 12\\
19& SWIRE  & 104605.56+583742.4		& 2.12   & 1.2	 & 45.70& 13\\
20& SWIRE& 104627.26+592526.9		& 0.67   & 1.6& 44.85& 6\\
21& SWIRE& 104652.52+571154.9		& 1.0   &  4.9& 45.69& 6\\
22&  SWIRE  & 104847.15+572337.6		& 1.47	& 4.5 & 45.98& 8\\
23& SWIRE& 105359.22+592103.0		& 0.95   & 2.6& 45.37& 6\\
24 & SWIRE & 105404.32+563845.6		& 1.72   &  6.5	 & 46.27& 8\\
25& IRAS 11095-0238  & 111203.4+020422	& 0.106	& 83 & 44.90& 2\\
26& IRAS 11130-2700 &  111531.6-271623 	& 0.136	& 25 & 44.60& 2\\ 
27& FSC 11257+5113 & 112832.7+505721	& 0.197	& 10.5 & 44.56& 3\\
28 &IRAS 12018+1941 &120424.5+192510 &0.169 &43 &45.03 &1\\
29 &IRAS 12071-0444 &120945.1-050113 &0.128 & 66& 44.97&1\\
30& IRAS 12127-1412  & 121518.9-142945	& 0.133	& 72 & 45.04& 2\\
31& NGC4388 & 122546.7+123944	& 		0.0084& 	2300 & 44.13& 4\\
32 &IRAS 12514+1027 &125400.8+101112 &0.319 & 64 &45.78&1\\
33 &Mrk 231 &125614.2+565225.2 &0.042 &1380&45.31&1\\
34 &  FSC 13297+4907 & 133150.5+485150	& 0.128	& 11.5 & 44.21& 3\\
35&Mrk 273 &134442.1+555312.7 &0.038 &287&44.53&1\\
36 &IRAS 14070+0525 &140931.3+051131 &0.264 &29&45.26&1\\
37& NGC5506& 141314.8-031227 	& 0.0062	&2000 & 43.79& 4\\
38& 	SST24 & 142648.90+332927.2		& 1.82   & 2.6 	 & 45.91& 13\\
39& SST24 & 142731.37+324434.6		& 0.90    & 1.8	 & 45.16 & 7\\
40&  SST24 & 142924.83+353320.3	& 2.73   & 1.5	& 45.98& 10\\
41&  SST24&  142958.33+322615.4	& 2.64   & 1.3	& 46.02& 10\\
42& SST24 & 143001.91+334538.4		& 2.46   & 5.5  & 46.47& 13\\
43& SST24 & 143050.84+344848.5			& 1.21    & 5.0	 & 45.86& 9\\
44& SST24&  143205.63+325835.2			& 0.48  &  5.7	& 45.10& 5\\
45&  SST24  & 143234.92+333637.5		& 1.12    & 5.0	 & 45.80 & 7\\  
46&  SST24 & 143301.49+342341.5		& 2.10    & 2.0	 & 45.95& 7\\
47& SST24  & 143341.90+330136.9			& 0.81   &  2.7	 & 45.25& 9\\
48& SST24 & 143508.49+334739.8		& 2.00   & 3.2 	 & 46.08& 13\\
49 & SST24 & 143520.75+340418.2 & 2.08  & 2.2 & 45.95& 10\\ 
50&  SST24 & 143523.99+330706.8	& 2.59   & 1.6 & 45.97& 10\\
51&   SST24 & 143539.34+334159.1	& 2.52  &  3.6 & 46.31& 10\\
52& SST24 & 143644.22+350627.4		& 1.77   & 3.0  & 45.95& 13\\
53&  SST24  & 143830.62+344412.0			& 0.94   &  3.6	 & 45.50& 9\\
54&IRAS 14378-3651 &144059.0-370432 &0.068& 85&44.69&1\\
55&  FSC 14448-0141 & 144727.5-015330	& 0.210	& 11.7	 & 44.66& 3\\
 56&  FSC 14475+1418 & 144954.9+140610	& 0.251	& 34.8	 & 45.30& 3\\
 57& FSC 14503+6006 & 145135.0+595437	& 0.577	& 22.4 & 45.86& 3\\
58& FSC 15065+3852 & 150825.4+384122	& 0.355	& 13.3 & 45.19& 3\\
59& IRAS 15225+2350 & 152443.9+234011 	& 0.139	& 43 & 44.86& 2\\
60 &Arp 220 &153457.1+233011 &0.018 &500&44.13&1\\
61 & FSC 15492+3454 & 155108.9+344533	& 0.311	& 13.6 & 45.08& 3\\
 62& FSC 15496+0331 & 155206.2+032244	& 0.193	& 28.9 & 44.98& 3\\
63& SWIRE & 160532.69+535226.4		& 2.75   & 4.1 & 46.43 & 13\\
64 &  FSC 16073+0209 & 160949.7+020130	& 0.223	& 38.7 & 45.24& 3\\
65&IRAS 16090-0139 &161140.5-014705 &0.134&74&45.06&1\\
66 &  FSC 16156+0146 & 161809.4+013922	& 0.133	& 72.8 & 45.05& 3\\
 67&  FSC 16242+2218 & 162626.0+221145	& 0.157	& 6.4 & 44.14& 3\\		
68& SWIRE & 164216.93+410127.8		& 2.40   & 4.2	 & 46.34& 13\\
69& IRAS 17044+6720 & 170428.4+671623 	& 0.135	& 53 & 44.92 & 2\\
70& SST24& 170916.91+585220.5		& 0.24   & 1.8	& 43.97& 6\\
71&  SST24 &  170939.20+592728.3 	&  2.52   &  2.3&  46.11&  11\\ 
72&  SST24 & 171057.45+600745.2	&  2.40   &  2.9&  46.18&  11\\ 
73& SST24& 171115.23+594907.1		&0.59     &9.9	&45.53& 6\\
74&  SST24 &  171144.26+585225.8 	&  1.22  &   1.4 &  45.32&11\\   
75& SST24& 171302.37+593611.0		& 0.684  & 7.4& 45.53& 5\\
76& AMS02& 171315.88+600234.2		& 1.8	& 0.55	 & 45.23 &15\\
77& AMS04&171340.62+594917.1		& 1.78   & 1.5	 & 45.66 &15\\
78& AMS06&171343.91+595714.6		& 1.8	& 1.9	 & 45.77 &15\\
79& AMS08&171429.67+593233.5		& 1.98   & 1.6	 & 45.77 &15\\
80&  SST24  & 171433.17+593911.2		& 2.45   & 1.95   & 46.02 & 14\\
81& SST24  & 171433.68+592119.3		& 1.00  &  0.7	 & 44.84 & 14\\	
 82& AMS09&	171434.87+585646.4		& 2.1    & 1.0	 & 45.61 &15\\
83&  SST24  & 171439.57+585632.1		& 1.85   & 1.35	 & 45.64 & 14\\	
84&  SST24  & 171510.28+600955.2		& 2.60   & 2.8	 & 46.22 & 14\\	
85&   SST24  & 171535.78+602825.5		& 2.48   & 1.5   & 45.91 & 14\\		
86&   SST24  & 171536.34+593614.8		& 2.70   & 1.8	 & 46.06 & 14\\	
87& 	SST24  & 171538.18+592540.1		& 2.65   & 3.7	 & 46.36 & 14\\		 
88&   SST24  & 171543.54+583531.2		& 2.47  &  2.6    & 46.15 & 14\\
89&  	SST24  & 171611.81+591213.3		& 2.15   & 1.4	 & 45.78 & 14\\	
90& SST24& 171750.65+584745.3		& 2.43   & 6.2	& 46.52& 6\\  
91&  SST24  & 171758.44+592816.8 		& 1.95  & 5.1	 & 46.26& 14\\	
92&   SST24  & 171826.67+584242.1		& 1.77   & 2.5	 & 45.87 & 14\\	
93&   SST24  & 171844.38+592000.5		& 2.08   & 4.5	 & 46.26 & 14\\	
94&   SST24  & 171844.77+600115.9		& 2.04   & 1.5	 & 45.76 & 14\\	
95& AMS14&	171845.47+585122.5		& 1.79   & 1.3	 & 45.60 &15\\
96&   SST24  & 171848.80+585115.1		& 2.34  &  2.3	 & 46.06 & 14\\	
97& SST24&  171852.71+591432.0			& 0.329  & 12.7	& 45.11& 5\\
98& AMS15&	171856.93+590325.0		& 2.1    & 0.6	 & 45.39 &15\\
99&  SST24  & 172045.17+585221.4		& 3.20   & 3.2   & 46.43 & 14\\	
100& AMS18	& 172046.32+600229.6		& 1.6    & 2.0	 & 45.70 &15\\
101&  SST24  & 172047.47+590815.1		& 2.59   & 1.1	 & 45.81 & 14\\	
102&  SST24 &  172048.02+594320.6	&  2.29   &  2.2 &  46.02&  11\\ 
103&   SST24  & 172051.48+600149.1		& 1.97   & 1.3	 & 45.68 & 14\\	
104&   SST24  & 172100.39+585931.0		& 1.84   & 1.6	 & 45.71 & 14\\	
105&   SST24  & 172119.46+595817.2		& 1.64   & 3.0	 & 45.89 & 14\\	 
106&   SST24 & 172126.42+601646.1		& 1.90   & 1.5 & 45.71 & 14\\	
107&   SST24  & 172422.10+593150.8		& 2.13   & 1.8	 & 45.88 & 14\\	
108&   SST24  & 172428.44+601533.2		& 2.43   & 1.65	 & 45.94 & 14\\	
109&  SST24 &  172448.65+601439.1	&  2.40   &  3.0&  46.19&  11\\ 
110&  FSC 17233+3712 & 172507.40+370932.1	& 0.702	& 8.1 & 45.60& 3\\
111& SST24 & 172542.34+595317.5		& 0.43   & 2.6 & 44.66 & 6\\ 
112 &IRAS 20100-4156 &201329.5-414734 &0.130 & 77&45.05&1\\
113 &IRAS 20551-4250 &205826.8-423900 &0.043 &269&44.62&1\\
114& NGC7172	& 220201.9-315211	& 	0.0087& 	2500 & 44.19& 4\\
115 &IRAS 23498+2423 &235226.0+244016 &0.212 &33&45.13&1\\

\enddata
\tablenotetext{a}{Observed flux density in mJy at 8 \um continuum peak measured from published spectra or from our own extractions, as described in section 2.3.  Typical uncertainties are $\pm$ 5\% for sources with f$_{\nu}$(8 $\mu$m) $\ga$ 5 mJy and $\pm$ 10\% for sources with f$_{\nu}$(8 $\mu$m) $\sim$ 1 mJy. }
\tablenotetext{b}{Rest frame luminosity $\nu$L$_{\nu}$(8 $\mu$m) in ergs s$^{-1}$ determined using peak f$_{\nu}$(8 $\mu$m) and luminosity distances from E.L. Wright, http://www.astro.ucla.edu/~wright/CosmoCalc.html, for H$_0$ = 71 \kmsMpc, $\Omega_{M}$=0.27 and $\Omega_{\Lambda}$=0.73. (Log [$\nu$L$_{\nu}$(8 $\mu$m)(\ldot)] = log [$\nu$L$_{\nu}$(8 $\mu$m)(ergs s$^{-1}$)] - 33.59.)}
\tablenotetext{c}{1 = \citet{far07}; 2 = \citet{ima07}; 3 = \citet{sar08}; 4 = \citet{shi07}; 5 = \citet{wee09}; 6 = new spectra in Figure 2 [sources 73,111 from archival program 30447 (G. Fazio, P.I.), source 90 from archival program 20629 (L. Yan, P.I.), source 16 from archival program 40539 (G. Helou, P.I), sources 13,20,21,23,70 from our own new observations, program 50031]; 7 = \citet{brn08a}; 8 = \citet{far09}; 9 = \citet{brn08b} ; 10 = \citet{hou05} ; 11 = \citet{wee06a} ; 12 = \citet{wee06c} ; 13 = \citet{pol08} ; 14 = \citet{yan07,saj08} ; 15 = \citet{mar08}. }
\end{deluxetable}

\clearpage

\begin{deluxetable}{cccccccc} 
\tablecolumns{8}
\tabletypesize{\footnotesize}

\tablewidth{0pc}
\tablecaption{Continuum Luminosities for Silicate Emission AGN}
\tablehead{
 \colhead{Source} &\colhead{Name} & \colhead{J2000 coordinates}&\colhead{z} & \colhead{f$_{\nu}$(8 $\mu$m)\tablenotemark{a}}& \colhead{log $\nu$L$_{\nu}$(8\,\um)\tablenotemark{b}}&\colhead{log L$_{ir}$\tablenotemark{c}} & \colhead{Ref.\tablenotemark{d}} 
}
\startdata

1 & I Zw 1  & 	005334.94+124136.2 & 0.061	 & 270 	& 44.93 &\nodata & 16\\
2 & PG0157+001  & 015950.21+002340.6 &  0.1630 & 36 &  44.93	& 46.38	& 17\\ 
3 & SWIRE  &  021640.72-044405.1 & 0.870\tablenotemark{e} & 9.5  & 45.86 &\nodata & 20\\
4 & SWIRE & 021729.06-041937.8	& 1.146\tablenotemark{e} & 4.6 	& 45.78	& \nodata	& 20\\
5 & SWIRE & 021743.04-043625.1 	& 0.784\tablenotemark{f} 	& 3.0 	&  45.26	& 	& 20\\
6 & SWIRE	& 021808.22-045845.3 	& 0.712 \tablenotemark{e}	&  5.2 	& 45.42	& \nodata	& 20\\
7 & SWIRE &021830.57-045622.9		& 1.401\tablenotemark{e} 	& 6.2		&  46.08	& \nodata	& 20\\
8	& SWIRE	& 021938.70-032508.2 	& 0.435\tablenotemark{e} 	&  2.1	&  44.58	& \nodata	& 20\\
9 & SWIRE	& 022012.21-034111.8\tablenotemark{e}		&  0.166 	&  2.7		&  43.82	& \nodata	& 20\\
10 &SWIRE  &  022431.58-052818.8  & 2.068\tablenotemark{e} & 9.7  & 46.59&\nodata & 20\\
11 	& IRAS07598+6508 &  080433.1+645948.6  & 0.148 & 200  & 45.58 & 46.14  	& 1\\
12 & PG 0804+761	 & 081058.60+760242.0 & 0.100	& 85 & 44.86 &\nodata & 16\\
13 & PG0838+770  & 084445.26+765309.5 & 0.1310  & 13.6  &  44.31 & 45.16& 17\\	
14 & PG1001+054  & 100420.13+051300.4& 0.1605  & 18.8  & 44.64	& 45.23	& 17\\ 
15	&  SWIRE & 103724.79+580513.5 & 1.5159\tablenotemark{g} & 6.1  &  46.14&\nodata & 20\\
16	& SWIRE  & 103803.38+572701.4  & 1.285\tablenotemark{g}  &  11.2  & 46.26&\nodata & 19\\
17	& SWIRE  &104255.68+575550.2 & 1.4684\tablenotemark{g} &  5.0  &  46.02&\nodata & 20\\
18	& SWIRE  &104705.12+590728.5  & 0.3913\tablenotemark{g}  &  3.2   &  44.66&\nodata & 20\\
19	& SWIRE  &105106.20+591625.1  & 0.7676\tablenotemark{g}   & 4.0   &  45.37&\nodata & 20\\
20	& SWIRE  &105158.57+590651.5  & 1.8131\tablenotemark{g}  & 5.7   &  46.25&\nodata & 20\\
21	& SWIRE  &105404.14+574019.8  & 1.1021\tablenotemark{g}  & 5.2   &  45.80&\nodata & 20\\
22	& SWIRE  &105959.96+574847.8  & 0.4529\tablenotemark{g}  & 3.7   &  44.86&\nodata & 20\\
23	& SWIRE  &110223.65+574435.9  & 0.2262\tablenotemark{g}  & 3.3   &  44.18	&\nodata  & 20\\
24 & PG1126-041  & 112916.66-042407.6 &  0.0600  & 41.7  &  44.11& 	 44.96 & 17\\  
25 &  PG 1211+143  & 121417.70+140312.6& 0.081 & 106	 & 44.77 &\nodata & 16\\
26 & 3C 273  & 	122906.70+020308.6 & 0.158	& 220	& 45.69	&46.41  & 16\\ 
27 & PG1229+204  &  123203.60+200929.2 &  0.0630  & 31.6 &   44.03 & 	44.96& 17\\	
28 & PG1244+026  & 124635.25+022208.8& 0.0482  & 17.7 &  43.55& 	44.50	& 17\\ 
29 & PG1302-102  & 130533.01-103319.4&  0.2784  & 23.9 & 45.24 & 	46.17& 17\\  
30 &  PG1351+640  & 135315.83+634545.6& 0.088	& 66 & 44.64&\nodata & 16\\			 
31 & PG1411+442  & 141348.33+440013.9& 0.0896  & 68.8 & 44.68 	& 44.96	 & 17\\
32&  SST24  & 142614.87+350616.5    &0.217   & 3.0   & 44.10&\nodata & 5\\
33 & PG1426+015  & 142906.59+011706.5& 0.0865  & 61.7 & 	44.60 	& 45.16	 & 17\\
34 &   SST24  & 143132.17+341417.9   &1.037  & 7.0   & 45.88&\nodata & 5\\
35  &   SST24  & 143156.40+325138.1 &   0.412  &3.1  &   44.70&\nodata & 5\\
36 &   SST24  & 143157.96+341650.1   &0.715  &7.8  &  45.60&\nodata & 5\\
37 & SST24  & 143310.33+334604.5	& 	2.4 & 12.7	& 46.82&\nodata & 5\\
38&   SST24  & 143409.54+334649.4 &0.216  & 3.1   & 44.11&\nodata & 5\\
39 & PG1440+356   & 144207.46+352622.9& 0.0791  & 61.7 & 44.52& 	45.18 & 17\\
40	& SWIRE & 155936.14+544203.6  & 0.3077\tablenotemark{g}  & 4.4   &  44.59&\nodata & 20\\
41	& SWIRE & 160128.58+544521.3  & 0.7277\tablenotemark{g}  &  6.9  &  45.56&\nodata & 20\\
42	& SWIRE & 160655.38+534016.9  & 0.2137\tablenotemark{g}  &  3.9   &  44.20&\nodata & 20\\
43 & PG1613+658  &  161357.18+654309.6&  0.1290  & 62.4 & 44.96 & 45.65 & 17\\
44	& SWIRE & 163111.31+404805.2  & 0.2575\tablenotemark{g}  &  6.8   &  44.61&\nodata & 20\\
45 & PG1700+518 &170124.80+514920.0& 0.2920  & 57.6 & 	45.66  & 46.37& 17\\ 
46	&  SST24 & 171124.25+593121.7  & 1.4904\tablenotemark{g}  &  4.2   &  45.96&\nodata & 20\\
47	& SST24	& 171233.48+583610.5	& 1.663\tablenotemark{e}	&  3.4	&  45.96	& \nodata	& 20\\
48 &SST24 & 171313.96+603146.6   &0.105  &  5.1 & 43.68&\nodata & 5\\
49 &SST24 & 171352.41+584201.2   &0.521\tablenotemark{e}  & 9.6  & 45.40&\nodata & 5\\
50	& SST24	& 171430.70+584225.0	&  0.562\tablenotemark{e} 	& 3.6		& 45.04	& \nodata	& 20\\
51	& SST24 &  171708.67+591341.1   &  0.645\tablenotemark{e}  &  2.8   & 45.06&\nodata & 19\\
52	& SST24 &171747.58+593258.0	  & 0.2477\tablenotemark{g}  &  2.9   &  44.21&\nodata & 20\\
53  &SST24 & 171839.73+593359.6    &0.382\tablenotemark{e}  & 4.1  & 44.75&\nodata & 5\\
54  & SST24 & 171902.29+593715.9   &0.178\tablenotemark{e}   & 9.1  & 44.40&\nodata & 5\\
55	& SST24 &172238.73+585107.0   & 1.617\tablenotemark{g}  &  5.4   & 46.14&\nodata & 20\\
56	& SST24 &172328.40+592947.3   & 1.34\tablenotemark{e}  &  4.8  & 	45.93&\nodata & 19\\
57	& SST24 & 172619.84+601600.2   & 0.925\tablenotemark{g}  &  3.7  & 45.50&\nodata & 20\\
58 & SST24 & 172704.67+593736.6	&  1.13	& 14.7	 & 46.27&\nodata & 5\\
59	&  PG2112+059	& 211452.57+060742.5 & 0.466	&   40   & 45.92 &\nodata   & 18\\
60 & PG2349-014 & 235156.12-010913.3& 0.1740  & 23.7 & 44.81	& 45.61 & 17\\

\enddata
\tablenotetext{a}{Observed flux density in mJy in continuum at rest frame 8 $\mu$m measured from published spectra in references given.  Typical uncertainties are $\pm$ 5\%.  For sources in \citet{sch07} (reference 17), f$_{\nu}$(8 $\mu$m) is determined by scaling from published f$_{\nu}$(6 $\mu$m) using the average spectra in Figure 2 of Schweitzer et al. for sources without PAH (characteristic of the highest luminosity sources), such that f$_{\nu}$(8 $\mu$m)/f$_{\nu}$(6 $\mu$m) = 1.12. }
\tablenotetext{b}{Rest frame luminosity $\nu$L$_{\nu}$(8 $\mu$m) in ergs s$^{-1}$ determined using luminosity distances D$_{L}$ from E.L. Wright, http://www.astro.ucla.edu/~wright/CosmoCalc.html, for H$_0$ = 71 \kmsMpc, $\Omega_{M}$=0.27 and $\Omega_{\Lambda}$=0.73. (Log [$\nu$L$_{\nu}$(8 $\mu$m)(\ldot)] = log [$\nu$L$_{\nu}$(8 $\mu$m)(ergs s$^{-1}$)] - 33.59.)}
\tablenotetext{c}{For sources in \citet{sch07} (reference 17), L$_{ir}$ is determined as in \citet{san96} for sources having measured IRAS flux densities at all IRAS wavelengths; L$_{ir}$  =  2.14x10$^{36}$ D$_{L}$$^{2}$[13.5(f$_{\nu}$(12 $\mu$m)) + 5.2(f$_{\nu}$(25 $\mu$m)) + 2.58(f$_{\nu}$(60 $\mu$m)) +(f$_{\nu}$(100 $\mu$m))] for flux densities in mJy and luminosities in ergs s$^{-1}$  (\ldot = 3.83x10$^{33}$ ergs s$^{-1}$).}  
\tablenotetext{d}{References for IRS spectrum:  1 = \citet{far07}; 5 = \citet{wee09}; 16 = \citet{hao05}; 17 = \citet{sch07}; 18 = \citet{mk07}; 19 = new extractions of sources in the FLS and SWIRE Lockman fields having f$_{\nu}$(24 $\mu$m) $>$ 5 mJy and $R$ $>$ 20 mag. which have optical redshifts and classifications cited as type 1 AGN in NED; 20 = new extractions of sources available within $Spitzer$ Legacy Program 40539 (G. Helou, P.I.), a flux limited sample with f$_{\nu}$(24 $\mu$m) $>$ 5 mJy, which have optical redshifts and classifications as type 1 AGN cited in NED or as determined from our examination of SDSS spectra, not including 2 sources also in reference 19. } 
\tablenotetext{e} {Optical redshift and classification in \citet{lac07}.}
\tablenotetext{f} {Optical redshift and classification in \citet{sim06}.}
\tablenotetext{g} {Optical redshift and classification from SDSS.}

\end{deluxetable}

\clearpage
%
%

\begin{figure}
\figurenum{1}
\includegraphics[scale=0.90]{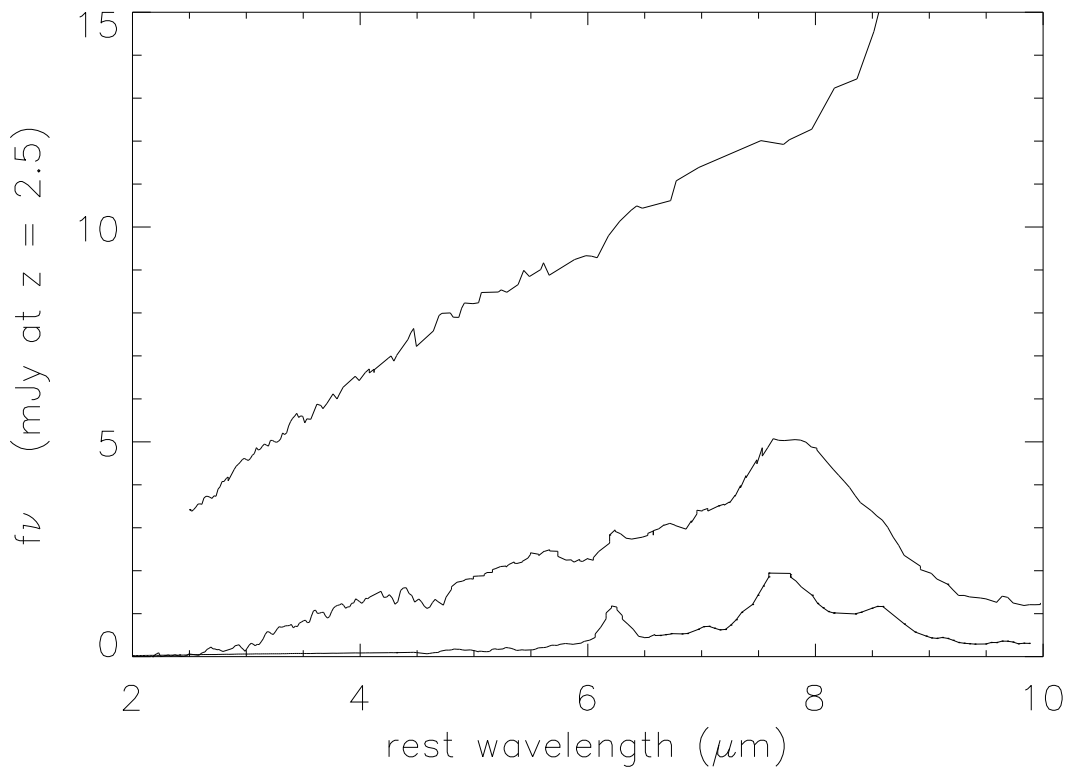}
\caption{Comparisons of the brightest sources yet discovered by $Spitzer$ and measured with the IRS at z $\sim$ 2.5.  Spectra are averages for the most luminous type 1 AGN (top), the most luminous obscured AGN (middle), and most luminous starburst (bottom).  All spectra are shown at the actual flux densities measured with the IRS at rest frame wavelengths for z = 2.5, although spectra are averages of brighter sources to improve displayed S/N, as described in section 2.1.1.}

\end{figure}

\begin{figure}
\figurenum{2}
 
\includegraphics[scale=0.9]{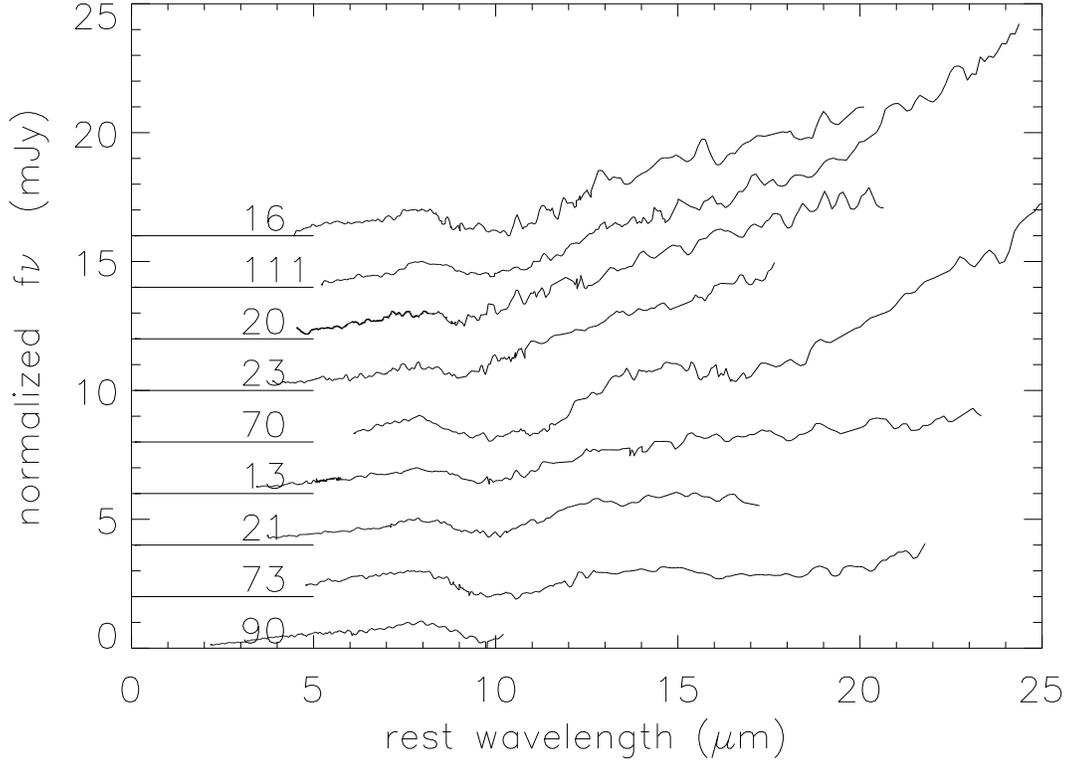}
\caption{Rest-frame spectra of all new absorbed sources in Table 1 which are taken from the SWIRE Lockman and First Look Survey fields because they have $R$ $>$ 20 and f$_{\nu}$(24 \ums) $>$ 5 mJy.  Of the 47 available IRS spectra of sources with these criteria, 11 show silicate absorption with optical depth of the 9.7 \um feature $\tau$ $>$ 0.7 and weak PAH emission, satisfying our definition of an absorbed AGN.  Nine of these spectra are shown above with source numbers from Table 1; sources 73, 111 are from archival program 30447 (G. Fazio, P.I.), source 90 from archival program 20629 (L. Yan, P.I.), source 16 from archival program 40539 (G. Helou, P.I.), and sources 13, 20, 21, 23, 70 from our own new observations, program 50031; The two previously published spectra are source 75 \citep{wee09} and source 17 in \citep{wee06c}. All spectra are normalized to f$_{\nu}$(8  \ums) = 1.0 mJy, but zero points are displaced for illustration.  The zero flux level for each spectrum is shown by the short horizontal line; scaling of actual flux densities of sources can be determined using observed values of f$_{\nu}$(8  \ums) given in Table 1. }  


\end{figure}

\begin{figure}
\figurenum{3}
\includegraphics[scale=0.80]{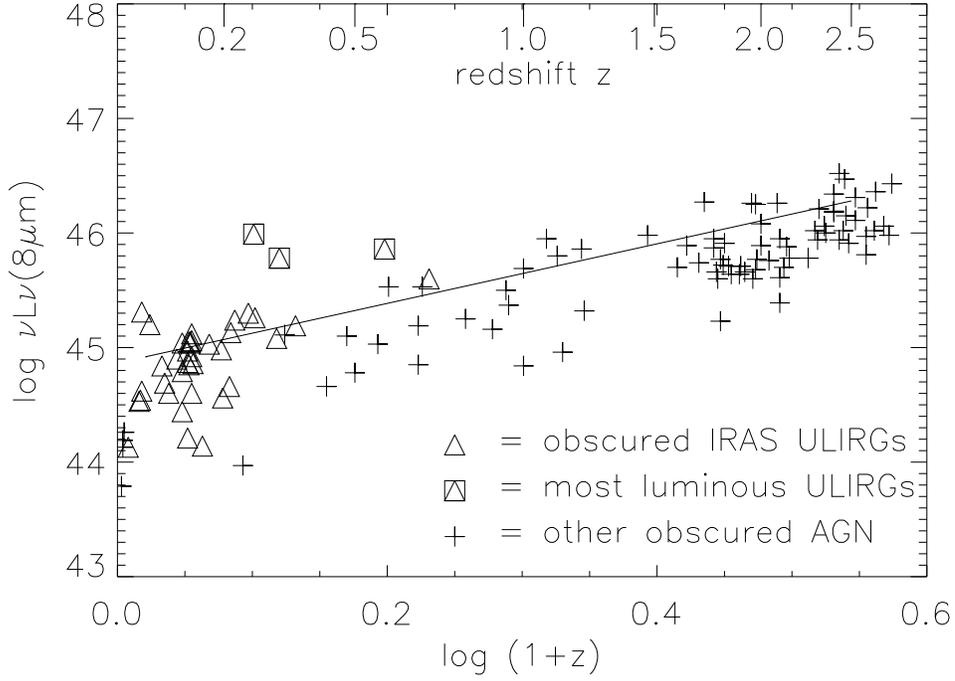}
\caption{AGN continuum luminosity $\nu$L$_{\nu}$(8 $\mu$m) in ergs s$^{-1}$ compared to redshift for all sources in Table 1 (obscured AGN). All luminosities are measured from sources observed spectroscopically with the IRS.  Triangles are IRAS ULIRGs with silicate absorption $\tau$ $>$ 0.7; crosses are other silicate absorbed AGN with $\tau$ $>$ 0.7. Solid line is luminosity evolution to z = 2.5 by factor (1+z)$^{2.6}$ determined from most luminous absorbed AGN, excluding the 3 most luminous ULIRGs shown as diamonds with squares (the 3 sources with spectra in Figure 5).  Luminosity uncertainties for ULIRGS arise primarily from flux calibration uncertainties and are typically $\pm$ 5\%; uncertainties for other sources, which have been discovered in $Spitzer$ MIPS surveys at 24 \um,  arise primarily from low signal to noise in observed spectra and are typically $\pm$ 10\%, which is smaller than plotted symbols. (Log [$\nu$L$_{\nu}$(8 $\mu$m)(\ldot)] = log [$\nu$L$_{\nu}$(8 $\mu$m)(ergs s$^{-1}$)] - 33.59.)} 

\end{figure}

\begin{figure}
\figurenum{4}
\includegraphics[scale=0.80]{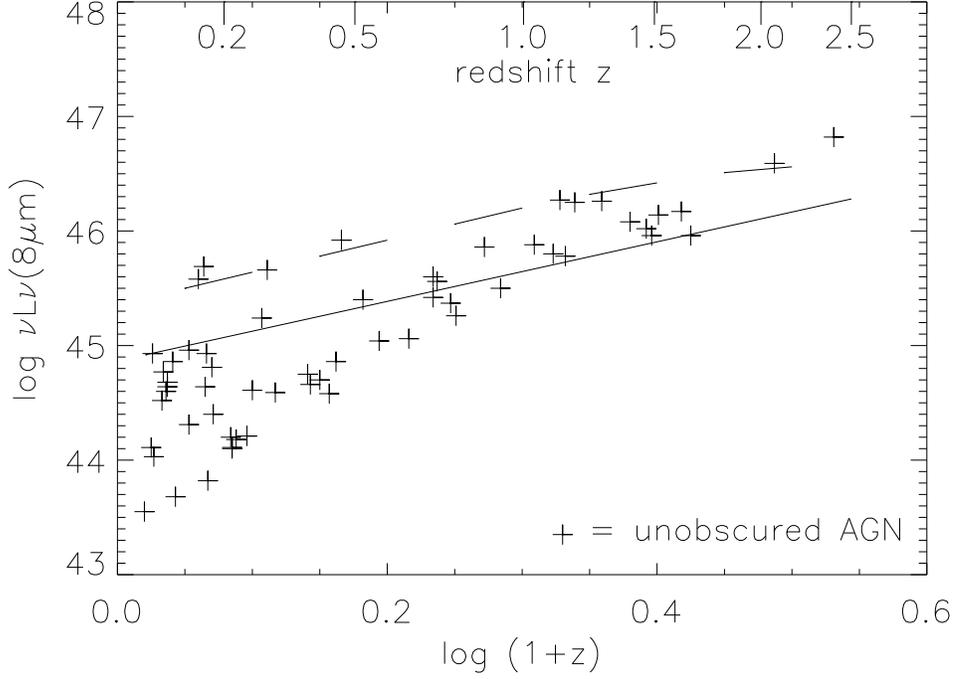}
\caption{AGN continuum luminosity $\nu$L$_{\nu}$(8 $\mu$m) in ergs s$^{-1}$ compared to redshift for all sources in Table 2 (unobscured, silicate emission AGN). All luminosities are measured from sources observed spectroscopically with the IRS.  Solid line is luminosity evolution to z = 2.5 by factor (1+z)$^{2.6}$ determined in Figure 3 for the most luminous obscured AGN. Long-dashed line is form of evolution for unobscured AGN determined by \citet{bro07} using a complete sample of Bootes sources with f$_{\nu}$(24 \ums) $>$ 1 mJy and optical redshifts and classifications as type 1; line is normalized to the brightest of these AGN observed with the IRS. Luminosity uncertainties arise primarily from low signal to noise in observed spectra and are typically $\pm$ 10\%, which is smaller than the plotted symbols. (Log [$\nu$L$_{\nu}$(8 $\mu$m)(\ldot)] = log [$\nu$L$_{\nu}$(8 $\mu$m)(ergs s$^{-1}$)] - 33.59.)} 

\end{figure}

\begin{figure}
\figurenum{5}
 
\includegraphics[scale=0.9]{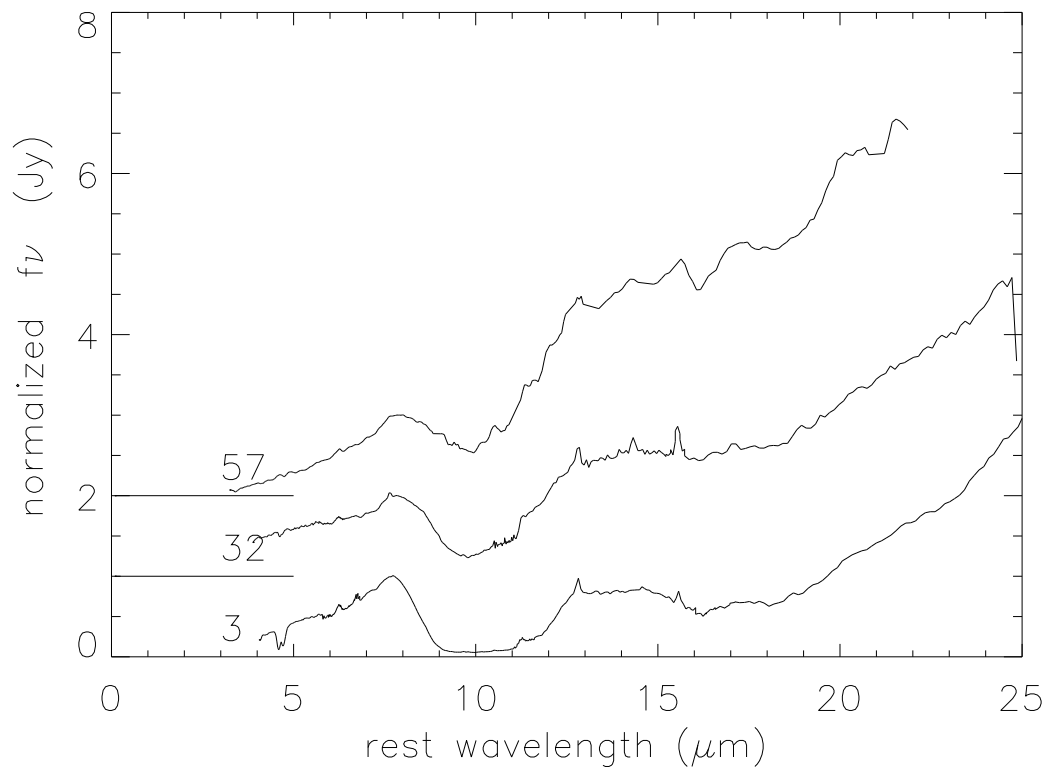}
\caption{Rest-frame spectra of the 3 most luminous ULIRGs in Table 1 and Figure 3 as determined by $\nu$L$_{\nu}$(8 $\mu$m).  Sources are labeled by the numbers in Table 1. All spectra are normalized to f$_{\nu}$(8 \ums) = 1.0 Jy, but zero points are displaced for illustration.  The zero flux level for each spectrum is shown by the short horizontal line; scaling of actual flux densities of sources can be determined using observed values of f$_{\nu}$(8 \ums) given in Table 1. }  

\end{figure}

\begin{figure}
\figurenum{6}
\includegraphics[scale=0.90]{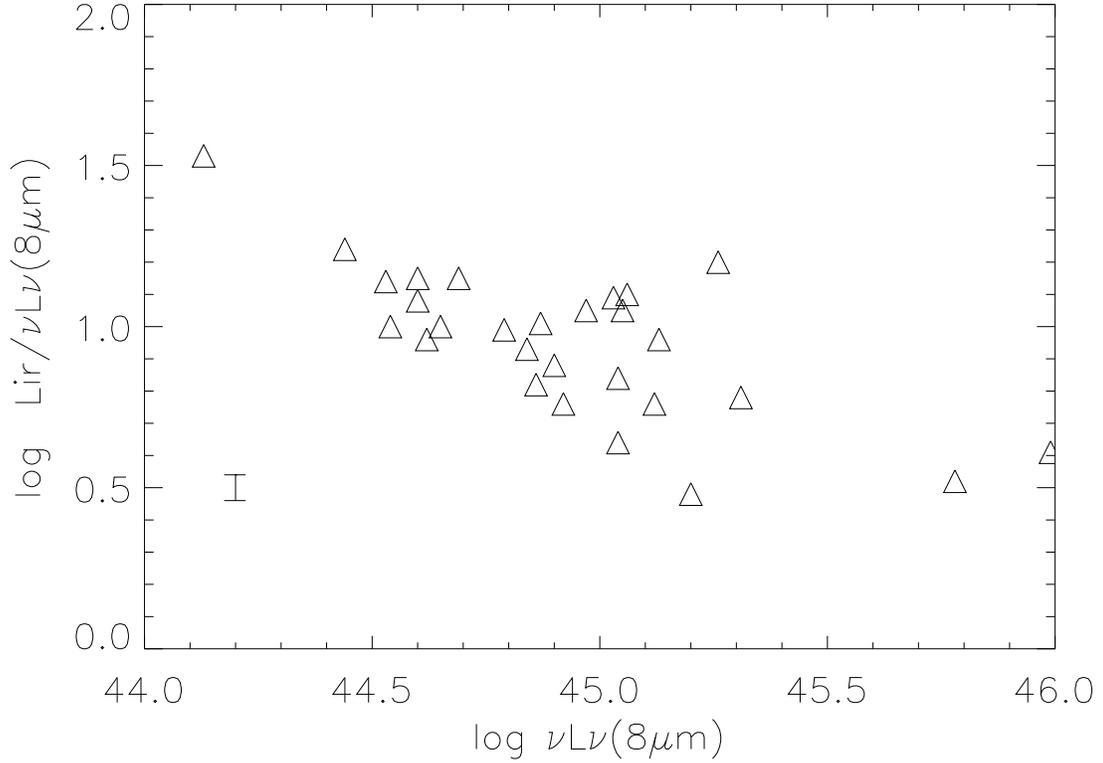}
\caption{Ratio of total infrared luminosity L$_{IR}$ to $\nu$L$_{\nu}$(8 $\mu$m) as function of $\nu$L$_{\nu}$(8 $\mu$m) for absorbed AGN in Table 1 which have total infrared luminosities measured from IRAS in \citet{far07} or \citet{ima07}. The two most luminous sources are sources 3 and 32 in Figure 5. Median ratio of 0.95$\pm$0.25 indicates that the typical correction for absorbed AGN is log L$_{IR}$ = log $\nu$L$_{\nu}$(8 $\mu$m) + 0.95$\pm$0.25 in ergs s$^{-1}$, or log L$_{IR}$ = log $\nu$L$_{\nu}$(8 $\mu$m) - 32.63$\pm$0.25, for L$_{IR}$ in \ldot~and $\nu$L$_{\nu}$(8 $\mu$m) in ergs s$^{-1}$ as used in Figure 3.  Error bar is uncertainty in ratio produced by observational uncertainty in $\nu$L$_{\nu}$(8 $\mu$m).} 

\end{figure}

\begin{figure}
\figurenum{7}
\includegraphics[scale=0.90]{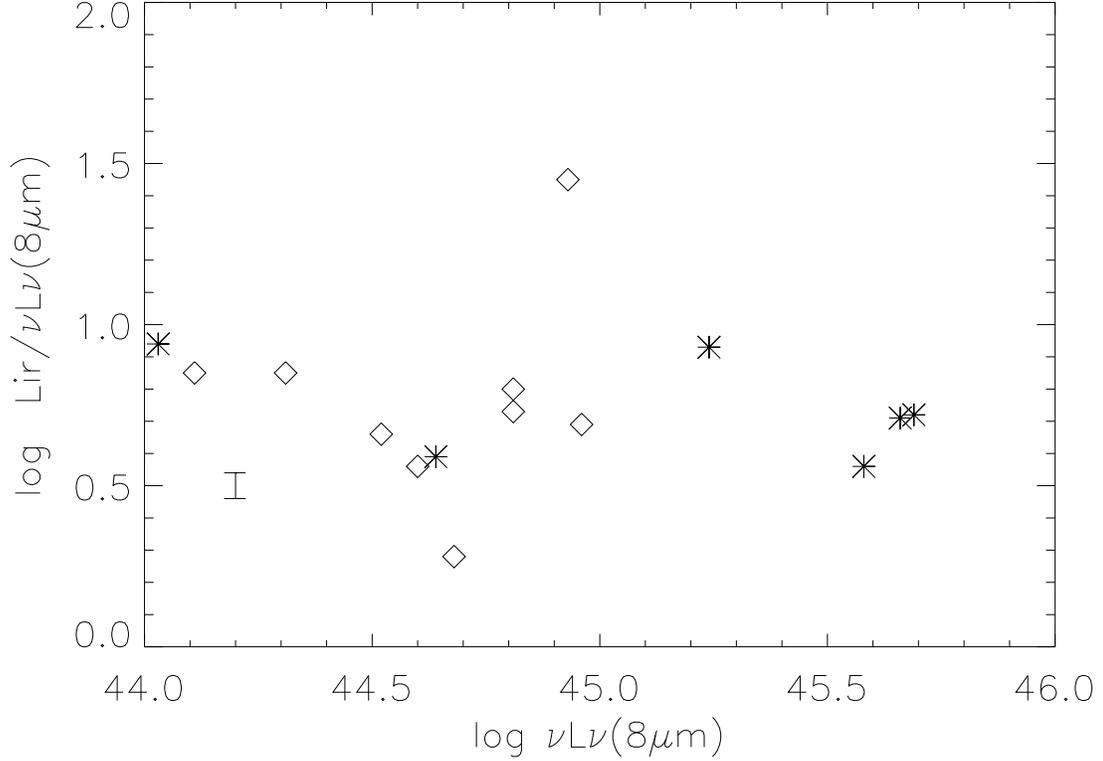}
\caption{Ratio of total infrared luminosity L$_{IR}$ to $\nu$L$_{\nu}$(8 $\mu$m) as function of $\nu$L$_{\nu}$(8 $\mu$m) for silicate emission, type 1 AGN in Table 2 which have total infrared luminosities measured at all IRAS wavelengths; source luminosities are in Table 2. The median ratio L$_{IR}$/$\nu$L$_{\nu}$(8 $\mu$m) = 0.74$\pm$0.15 indicates that the typical correction for unabsorbed type 1 AGN is log L$_{IR}$ = log $\nu$L$_{\nu}$(8 $\mu$m) + 0.74$\pm$0.15 in ergs s$^{-1}$, or log L$_{IR}$ = log $\nu$L$_{\nu}$(8 $\mu$m) - 32.84$\pm$0.15 for L$_{IR}$ in \ldot~and $\nu$L$_{\nu}$(8 $\mu$m) in ergs s$^{-1}$ as used in Figure 4.  Asterisks show values for sources without measured PAH features and diamonds for sources with weak PAH features.  Error bar is uncertainty in ratio produced by observational uncertainty in $\nu$L$_{\nu}$(8 $\mu$m).}

\end{figure}

\begin{figure}
\figurenum{8}
\includegraphics[scale=0.90]{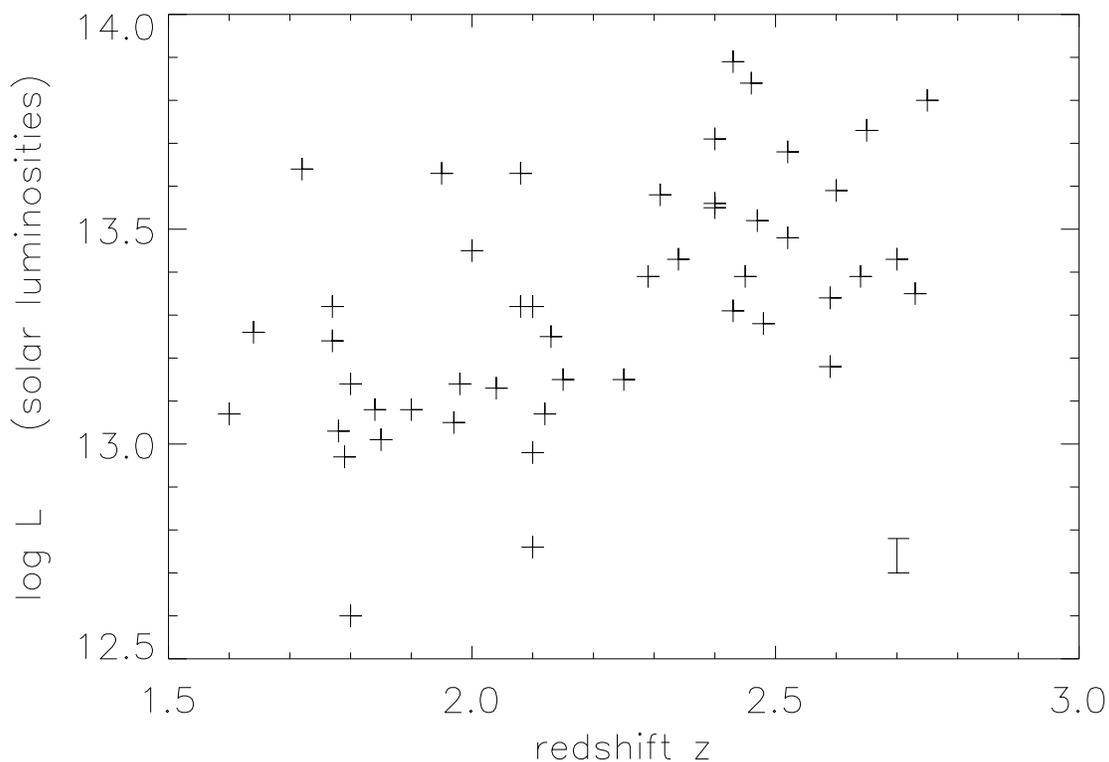}
\caption{Redshifts and luminosities for highest redshift obscured AGN.  Luminosities are  L$_{IR}$ in \ldot~using $\nu$L$_{\nu}$(8 $\mu$m) from Table 1 and transformation log L$_{IR}$ = log $\nu$L$_{\nu}$(8 $\mu$m) - 32.63$\pm$0.25, for L$_{IR}$ in \ldot~and $\nu$L$_{\nu}$(8 $\mu$m) in ergs s$^{-1}$.  Space densities and luminosity functions are determined for intervals 1.7 $<$ z $<$ 2.2 and 2.3 $<$ z $<$ 2.7 using distributions of points shown.  Error bar is uncertainty in ratio produced by observational uncertainty in $\nu$L$_{\nu}$(8 $\mu$m).} 

\end{figure}

\begin{figure}
\figurenum{9}
\includegraphics[scale=0.90]{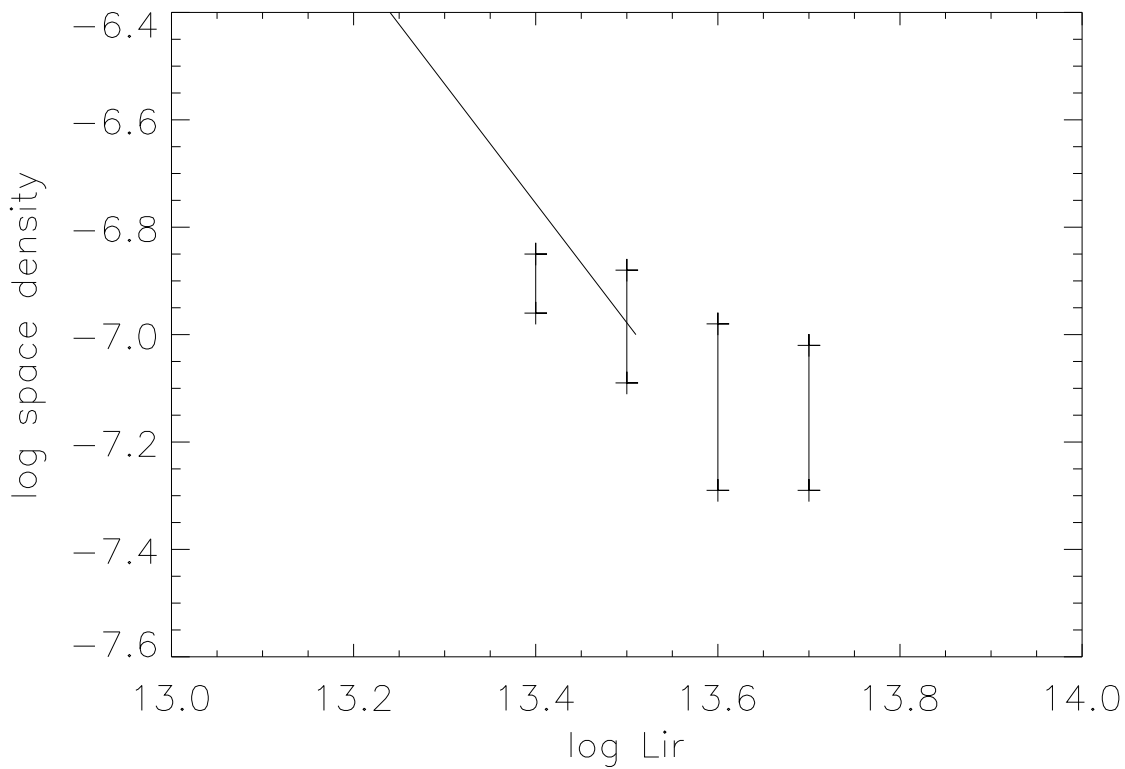}
\caption{Luminosity function at z = 2.5 determined from obscured AGN in Figure 8.  Space densities are in number Mpc$^{-3}$ per interval 0.4 in log L$_{IR}$, for L$_{IR}$ in \ldot.  Error bars show the differences in results for the two independent redshift windows 1.7 $<$ z $<$ 2.2 and 2.3 $<$ z $<$ 2.7.  Solid line is the bright end of the local luminosity function for all IRAS galaxies from \citet{soi87}, scaled to z = 2.5 using an 
evolution factor (1+z)$^{2.5}$.}
\end{figure}

\begin{figure}
\figurenum{10}
\includegraphics[scale=0.90]{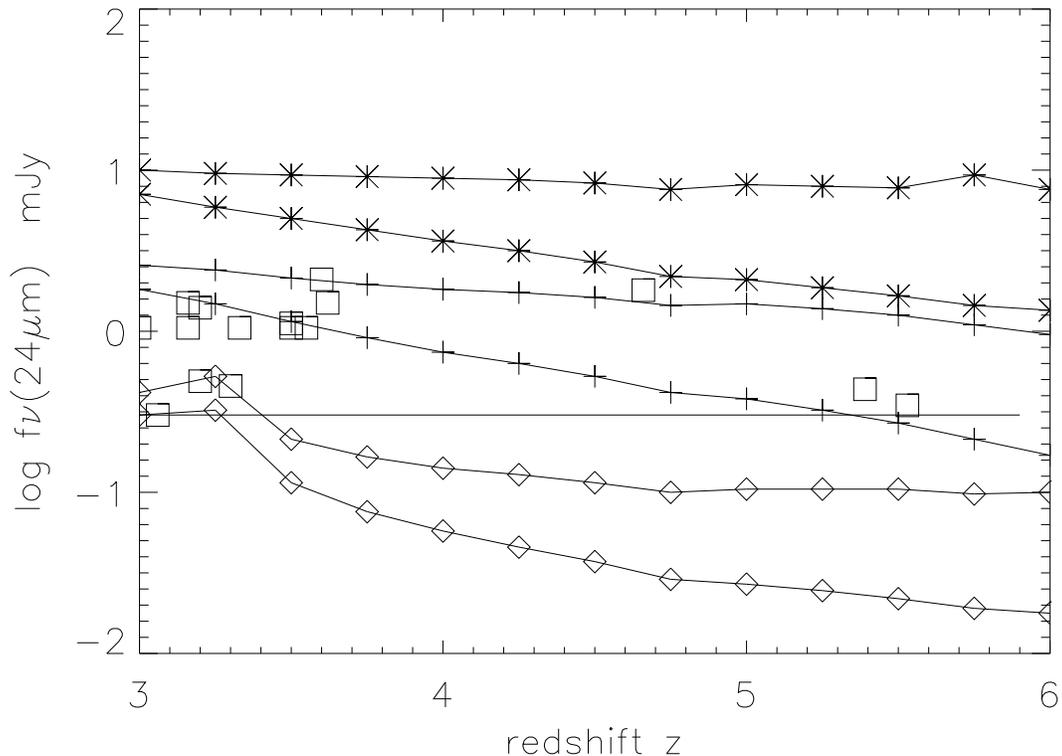}
\caption{Predicted fluxes measured with MIPS at 24 \um as a function of redshift for the most luminous sources known at z $\sim$ 2.5 when extrapolated to higher redshifts.  Asterisks are predictions for the most luminous, unobscured, type 1 AGN; crosses are predictions for the most luminous obscured AGN; and diamonds are predictions for the most luminous starbursts, all scaled to the brightest examples shown in Figure 1.  For each set, the upper curve shows expected f$_{\nu}$(24 $\mu$m) if source luminosities continue to scale as (1+z)$^{2.5}$ for 2.5 $<$ z $<$ 6, and lower curve shows expected f$_{\nu}$(24 $\mu$m) if there is no luminosity evolution for z $>$ 2.5.  Horizontal line is MIPS limit of f$_{\nu}$(24 $\mu$m) = 0.3 mJy for wide area surveys such as in Bootes, the FLS, and SWIRE. Squares are type 1 AGN discovered at 24 \um in $Spitzer$ surveys having optical classifications and redshifts, as cited in text.} 

\end{figure}

\end{document}